\documentclass[amsmath,amssymb,onecolumn]{IEEEtran}

 \usepackage{color}

\usepackage{graphicx}
\usepackage{amsfonts,amsmath,amssymb}
\usepackage{mathrsfs}

 \newcommand{\er}[1]{\hbox{(\ref{#1})}}

\newtheorem{theorem}            {Theorem}[section]
\newtheorem{corollary}          [theorem]{Corollary}

\newtheorem{definition}         [theorem]{Definition}
\newtheorem{lemma}              [theorem]{Lemma}
\newtheorem{sideremark}         [theorem]{Remark}
\newtheorem{sideeg}           [theorem]{Example}
\newtheorem{sideconj}           [theorem]{Conjecture}
\newtheorem{sideassumption}   [theorem]{Assumption}

\newenvironment{remark}         {\begin{sideremark}\rm}{\end{sideremark}}

\newenvironment{assumption} {\begin{sideassumption}\it}{\end{sideassumption}}

\def\tr{\mathop{\rm tr}\nolimits} 

\def\smhsp{\hspace{0.1cm}}
\def\trp{^T}
\def\cbb{\mathbb{C}}
\def\rbb{\mathbb{R}}
\def\diag{{\rm diag}}

\usepackage{enumerate,cite,latexsym,graphicx}

\begin{document}

\title{$H^\infty$  Control of Linear Quantum Stochastic Systems}
\author{Matthew R.~James\thanks{M.R. James is with the Department of
    Engineering, Australian
    National University, Canberra, ACT 0200,
    Australia. Matthew.James@anu.edu.au. Research supported by the
    Australian Research Council.}
    \and
    Hendra I.~Nurdin\thanks{H.I.~Nurdin is with the Department of
    Information Engineering, Australian
    National University, Canberra, ACT 0200,
    Australia. Hendra.Nurdin@anu.edu.au. Research supported by National ICT
    Australia, Ltd. (NICTA). National ICT Australia is funded by the
    Australian Government's Department of Communications, Information
    Technology and the Arts and the Australian Research Council
    through Backing Australia's Ability and the ICT Centre of
    Excellence Program.}
  \and
  Ian R.~Petersen\thanks{I.R. Petersen is with the School of
    Information Technology and Electrical Engineering, University of
    New South Wales at the Australian Defence Force Academy, Canberra,
    ACT 2600 Australia.  irp@routh.ee.adfa.edu.au. Research supported
    by the Australian Research Council.}}

\date{\today}

\maketitle


\begin{abstract}
The purpose of this paper is to  formulate and solve a $H^\infty$
controller synthesis problem for a class of  non-commutative
linear stochastic systems which includes many examples of interest
in quantum technology.  The paper includes results on the class of
such systems for which the quantum commutation relations are
preserved (such a requirement must be satisfied in a physical
 quantum system).
A quantum version
of standard (classical) dissipativity results are presented and from this a
quantum version of
the Strict Bounded Real Lemma is derived. This enables a quantum
version of the two Riccati solution to the $H^\infty$ control problem
to be presented. This result leads to  controllers  which may be realized using
purely quantum, purely classical or a mixture of  quantum and
classical elements. This issue of physical realizability of the controller is examined in detail, and necessary and sufficient conditions are given.
Our results are  constructive in the sense that we provide
explicit formulas for the Hamiltonian function and coupling operator
corresponding to the controller.

\end{abstract}


\section{Introduction}
\label{sec:introduction}

Recent developments in quantum and nano technology have provided a
great impetus for research in the area of quantum feedback control
systems; e.g., see \cite{VPB83,HW94a,DJ99,AASDM02,YK03a,YK03b,GSM04}. In particular, it is now being realized that
robustness is a critical issue in quantum feedback control systems,  as
it is in classical (i.e., non-quantum) feedback control systems; e.g., see \cite{ZR94,BR02,DJ06}.
However, the majority of feedback control results for quantum systems do not
address the issue of robustness directly. The aim of this paper is to address the problem of systematic robust
control system design for quantum systems via a $H^\infty$ approach.

We present  a $H^\infty$ controller synthesis result for a class
of non-commutative linear stochastic systems which includes many
examples of interest in quantum technology. The synthesis
objective is to find a disturbance attenuating controller which
bounds the influence of certain signals,
 called the disturbance input signals, on another set of signals, called the
performance output signals. In this way, the undesirable effects of disturbances on performance is reduced in a systematic and quantifiable way. This follows 
from a quantum version of
the small gain theorem \cite{DJ06}; indeed,  the controller will be
robustly stabilizing against certain kinds of uncertainties, which
in principle could include parameter uncertainties, modelling
errors, etc. To illustrate the results, we consider
some design examples in quantum optics. 

A feature of our approach
is that the control designer can choose to 
synthesize a controller which may be  quantum, classical or a
mixed quantum-classical controller for the plant. 
The majority of
the available results in quantum feedback control consider the
controller to be a classical (i.e., non-quantum) system, which may
be implemented using analog or digital electronics. Classical
controllers process measurement data obtained by monitoring the
quantum system to determine control actions which influence the
dynamics of the quantum system in a feedback loop. In contrast,
quantum controllers are themselves quantum systems, and the closed
loop is fully quantum; e.g., see
\cite{VPB91a,WM94b,SL00,YK03a,YK03b,DJ06}.
In \cite{YK03a,YK03b}, a transfer function approach to quantum
control based on the chain scattering approach to $H^{\infty}$
control has been proposed.  However, the plants and controllers
considered therein are SISO (single input single output) systems
having only quantum degrees of freedoms. Moreover, no systematic
treatment is given of the physical realizability of the resulting
controllers. On the other hand, our approach is developed for a
fairly general class of MIMO (multiple input multiple output)
quantum linear stochastic systems with possibly mixed quantum and
classical degrees of freedom and addresses the physical
realizability issue.

Our approach involves deriving a quantum version of the Strict Bounded
 Real Lemma (e.g., see  \cite{PAJ91}). We begin by considering a
 general problem of dissipativity for quantum systems in a manner
that generalizes Willems' theory of dissipative systems (see
\cite{WIL72}), originally developed for nonlinear deterministic
classical systems. The  paper  characterizes this dissipation
property in algebraic terms. This then leads to a quantum version
of the Strict Bounded Real Lemma. This lemma is then applied to
 the closed loop system formed from the interconnection between the
 quantum plant and the controller. By following
 an algebraic approach to the $H^\infty$ control problem such as in
 \cite{PAJ91}, this enables us to derive a quantum version of the
 celebrated two Riccati solution to the $H^\infty$ control problem;
 e.g., see \cite{ZDG96,GL95}.

 The two Riccati quantum $H^\infty$ result which is derived leads to formulas
 for some, but not all,  of  the controller state space matrices. Controller  noise sources (needed for physical realizability, as discussed shortly)  are not determined by these Riccati equations. If the designer chooses to synthesize a classical controller, then the standard classical $H^\infty$ controller suffices, and no further matrices nor noise sources need be determined.  However, if the designer chooses to synthesize a controller that is itself a quantum system, or contains a component that is a quantum system, then the controller design must be completed by 
selecting the undetermined matrices and noise sources  to ensure that the controller is physically meaningful.  
 For example, in a quantum controller,
 quantum mechanics dictates that the time evolution of a closed
 system preserve certain  commutation
 relations. This requirement
 constrains the possible controller matrices and noise sources  for a physically realizable controller.
 To address this issue, the paper considers the
 question of physical realizability. Starting with a standard
 parameterization of purely
 quantum linear systems in terms of a quadratic Hamiltonian
 and a linear coupling operator (e.g., see \cite{EB05}),  we then derive
 necessary and sufficient conditions for given controller state space
 matrices to be physically realizable. These conditions are
 constructive in that if a set of controller state space matrices are
 physically realizable, then we can construct the required Hamiltonian
 function and coupling operator.
 
 We begin in Section \ref{sec:models} by presenting the class of
 models under consideration and we present a result describing the
 condition such systems must satisfy in order to correspond to a  physical
quantum system in that the quantum commutation relations are
 preserved. In Section \ref{sec:perform-diss}, we consider the
 question of dissipation for
 quantum systems and derive a quantum version of the Strict Bounded
 Real Lemma.  In Section  \ref{sec:synth-H}, we
 set up the $H^\infty$ problem to be solved and present our main
 result which is
 a two Riccati solution to this quantum $H^\infty$ control
 problem. This section also considers the question of physical
 realizability.
In Section \ref{sec:appeg-robust}, we consider the
 application of our quantum $H^\infty$ control results to the question
 of stability robustness and we establish a version of the small gain
 theorem for quantum systems subject to parameter uncertainty.
 In Section
 \ref{sec:appeg}, we present some  examples from quantum optics  to
 illustrate the theory which has been developed. In particular, we
 consider the control of quantum optical plants using    quantum, classical, and quantum-classical controllers. We also consider the
 design of a purely quantum controller which leads to robustness
 against uncertainty in one of the physical parameters of the cavity
 system.
The paper is organized so that the appendix contains all of the proofs
of the results which are presented.

{\bf Acknowledgments.} The authors wish to thank V. P.~Belavkin,
E.~Huntington, H.~Mabuchi, J.~Gough, L.~Bouten and R.~van~Handel
for helpful discussions.


\section{Linear Quantum Stochastic Models}
\label{sec:models}

In this paper, we are interested in physical systems that contain one
or more components  that are quantum in nature. It is helpful to have
in mind an interconnection of components, some of which are
\lq\lq{classical}\rq\rq, meaning that non-quantum descriptions
suffice, and some for which   \lq\lq{quantum}\rq\rq \  descriptions
are required. Such systems are common in quantum optics laboratories,
and may occur, for instance, in schemes for implementing quantum
computing and information processing algorithms.
We use non-commutative or quantum probability theory  (e.g., see
\cite{BHJ06a} and the references therein) to describe the
systems of interest. This framework is quite general and encompasses
quantum and classical mechanical systems. Quantum noise, which may
arise from measurements or interactions between subsystems and the environment, is
central.

We consider linear non-commutative stochastic systems of the form
\begin{eqnarray}
dx(t) &=& Ax(t) dt + B dw(t); \quad x(0)=x_0
\nonumber \\
dy(t) &=& C x(t) dt + D dw(t) \label{linear-c}
\end{eqnarray}
where $A$, $B$, $C$ and  $D$   are, respectively, real $\rbb^{n
\times n},\rbb^{n \times n_w},\rbb^{n_y \times n}$ and $\rbb^{n_y
\times n_w}$ matrices ($n,n_w,n_y$ are positive integers), and $
x(t) = [\begin{array}{ccc} x_1(t) & \ldots & x_n(t)
\end{array}]\trp$ is a vector of self-adjoint possibly
non-commutative system variables.

The initial system variables $x(0)=x_0$ consist of operators (on
an appropriate Hilbert space) satisfying the {\em commutation
relations}\footnote{In the case of a single degree of freedom
quantum particle, $x=(x_1, x_2)^T$ where $x_1=q$ is the position
operator, and $x_2=p$ is the momentum operator. The annihilation
operator is $a=(q+ip)/2$. The commutation relations are
$[a,a^\ast]=1$, or $[q,p]=2 i$.}
\begin{equation}
[x_j(0), x_k(0) ] = 2 i \Theta_{jk}, \ \ j,k = 1, \ldots, n,
\label{x-ccr}
\end{equation}
where $\Theta$ is a real antisymmetric matrix with components
$\Theta_{jk}$, and $i=\sqrt{-1}$.  Here, the commutator is defined
by $[A,B]=AB-BA$. To simplify matters without loss of generality,
we take the matrix $\Theta$ to be of one of the following forms:
\begin{itemize}
\item \emph{Canonical} if $\Theta=\diag(J,J,\ldots,J)$, or

\item \emph{Degenerate canonical} if $\Theta=\diag(0_{n' \times
n'},J,\ldots,J)$, where $0< n' \leq n$.
\end{itemize}
Here, $J$ denotes the real skew-symmetric $2\times 2$ matrix
$$
J= \left[ \begin{array}{cc} 0 & 1 \\ -1 & 0
\end{array} \right],$$
and the \lq\lq{diag}\rq\rq \ notation indicates a block diagonal
matrix assembled from the given entries. To illustrate, the case
of a system with one classical variable and two conjugate quantum
variables is characterized by $\Theta=\mathrm{diag}(0,J)$, which
is degenerate canonical.  It is assumed that   $x_0$ is Gaussian, with density operator $\rho$.

The vector quantity $w$ describes the input signals and is assumed
to admit the decomposition
\begin{equation}
dw(t) = \beta_w(t)dt + d\tilde w(t) \label{w-general}
\end{equation}
where $\tilde w(t)$ is the noise part of $w(t)$ and $\beta_w(t)$
is a self adjoint, adapted process (see, e.g.,
\cite{HP84,KRP92,BHJ06a} for a discussion of adapted processes).
The noise $\tilde w(t)$ is a vector of self-adjoint quantum noises
with Ito table
\begin{equation}
d\tilde w(t) d \tilde w^T(t) = F_{\tilde w} dt,
\label{w-Ito-general}
\end{equation}
where $F_{\tilde w}$ is a non-negative Hermitian matrix; e.g., see
\cite{KRP92,VPB92}. This determines the following commutation
relations for the noise components:
\begin{equation}
[ d\tilde w(t), d\tilde w^T(t) ] = d\tilde w(t) d\tilde w^T(t) -
(d\tilde w(t) d\tilde w^T(t))^T =  2T_{\tilde w}dt , \label{w-ccr}
\end{equation}
where we use the notation $ S_{\tilde w} = \frac{1}{2} ( F_{\tilde
w} + F_{\tilde w}^T),  \ \ T_{\tilde w} = \frac{1}{2} ( F_{\tilde
w} - F_{\tilde w}^T)$ so that $F_{\tilde w} = S_{\tilde w} +
T_{\tilde w}$. For instance, $F_{\tilde w} =
\mathrm{diag}(1,I+iJ)$ describes a noise vector with one classical
component and a pair of conjugate quantum noises (here $I$ is the
$2\times 2$ identity matrix). The noise processes can be
represented as operators on an appropriate Fock space (a
particular, yet important, type of Hilbert space); e.g., see
\cite{KRP92,VPB92}.

The process $\beta_w(t)$ serves to represent variables of other
systems which may be passed to the system (\ref{linear-c}) via an interaction. Therefore, we require that $\beta_w(0)$ is an operator
on a Hilbert space  distinct from that of $x_0$ and the noise processes.
We also assume $\beta_w(t)$ commutes with $x(t)$
for all $t \geq 0$ (two vectors $x,y$ of operators are said to
commute if $xy\trp-(yx\trp)\trp=0$); this will simplify matters
for the present work. Moreover, since we had earlier specified
that $\beta_w(t)$ should be an adapted process, we make note that
$\beta_w(t)$ also commutes with $d\tilde w(t)$ for all $t \geq 0$. 

To simplify the exposition, we now set up some conventions to put
the system (\ref{linear-c}) into a standard form. First, note that
there will be no change to the dynamics of $x(t)$ and $y(t)$ if we
enlarge $w(t)$, by adding additional dummy noise components, and at the
same time enlarging $B$ by inserting suitable columns of zeros.
Secondly, we may add dummy components to $y$ by enlarging $C$ and
$D$ by inserting additional dummy rows to each of these matrices.
Our original output can be recovered by discarding or
``disconnecting'' the dummy components/entries. Therefore, we make
the following assumptions on the system (\ref{linear-c}): (i)
  $n_y$ is even, and (ii) $n_w \geq n_y$.
We also make an assumption that $F_{\tilde w}$ is of the
\emph{canonical} form $F_{\tilde w}=I+i\diag(J,\ldots,J)$. Hence
$n_w$ has to be even. Note that if $F_{\tilde w}$ is not canonical
but of the form $F_{\tilde w}=I+i\diag(0_{n' \times
n'},\diag(J,\ldots,J))$ with $n' \geq 1$, we may enlarge $w(t)$
(and hence also $\tilde w(t)$) and $B$ as before such that the
enlarged noise vector, say $\tilde w'$, can be taken to have an
Ito matrix $F_{\tilde w'}$ which is canonical.

Equation (\ref{linear-c}) is a linear quantum stochastic
differential equation. General quantum stochastic differential
equations of this type are described in \cite{HP84}, \cite{KRP92}, \cite{VPB92}. In  (\ref{linear-c})    the integral with respect to $dw(t)$ is taken to be a quantum
stochastic integral. The solution $x(t)$ depends only on the
past noise $w(s)$, for $0\leq s\leq t$; i.e., it is adapted, and a
property of the Ito increments is that $d\tilde w(t)$ commutes with
$x(t)$.

%

Equation \er{linear-c} describes a non-commutative linear
stochastic system, which need not necessarily correspond to a
  physical system. This issue does not normally arise in physical modeling, but as we shall see it is
 of considerable
importance when we come to synthesizing physically realizable
controllers below in Section \ref{sec:phys_real} and Subsection
\ref{sec:synth-H-real}. The following theorem
provides an algebraic characterization of precisely when the
linear system \er{linear-c} preserves the commutation relations as
time evolves, a property enjoyed by open physical systems undergoing an
overall unitary evolution, \cite{GZ00}. The proof is given in the appendix.

\begin{theorem} \label{thm:preserve}
Under the assumptions discussed above for the system
\er{linear-c}, we have $[x_i(0),x_j(0)] = 2i\Theta_{ij}$ implies
$[x_i(t),x_j(t)] = 2i\Theta_{ij}$ for all $t \geq 0$ if and only
if
\begin{equation}
iA \Theta + i\Theta A^T + B T_{\tilde w} B^T =0 \label{AB-ccr}
\end{equation}
\end{theorem}

\section{Physical realizability of linear QSDEs}
\label{sec:phys_real}

Unlike classical systems, which we may regard here as always being
physically realizable (for the purpose of controller synthesis),
at least approximately via classical analog or digital
electronics, a quantum system represented by the linear QSDE
(\ref{linear-c}) need not necessarily represent the dynamics of a
meaningful physical system. An example of a meaningful physical
system here could be a system made up of interconnection of
various quantum optical devices such as optical cavities, beam
splitters, optical amplifiers. In particular, we have already seen
from the previous section that in physical devices, the canonical
commutation relations need to be preserved for all positive times
leading to the requirement that the constraint (\ref{AB-ccr})  be
satisfied by the matrices $A$ and $B$ of (\ref{linear-c}).
However, as we shall shortly see, there is another constraint
related to the output signal $y(t)$ which is required for
(\ref{linear-c}) to be physically realizable.

\subsection{Open quantum harmonic oscillator}
\label{sec:qho}

%

In order to formally present a definition of an open quantum
harmonic oscillator we will require the following notation. For a
square matrix $T$, $\diag_{m}(T)$ denotes the block diagonal
matrix $\diag(T,\ldots,T)$ where $T$ appears $m$ times as a
diagonal block. The symbol $P_m $ denotes a $2 m \times 2 m$
\emph{permutation matrix} defined so that if we consider a column
vector $a=[
\begin{array}{cccc} a_1 & a_2
  & \ldots & a_{2
m}\end{array}]^T$, then $P_m a=[\begin{array}{cccccccc}a_1 & a_3
&\ldots & a_{2 m-1} & a_2 & a_4 & \ldots& a_{2m}\end{array}]\trp$.
An $m \times m$ permutation matrix is a full-rank real  matrix
whose columns (or, equivalently, rows) consist of  standard basis
vectors for $\rbb^{m}$; i.e., vectors in $\rbb^m$ whose elements
are all $0$ except for one element which has the value $1$. A
permutation matrix $P$ has the unitary property $PP\trp=P\trp P=
I$. Note that $P_m\trp [\begin{array}{cccc} a_1 & a_2 & \ldots &
a_{2m} \end{array}]\trp= [\begin{array}{cccccccc} a_1 & a_{m+1} &
a_2 & a_{m+2} & \ldots & a_{m} & a_{2m}\end{array} ]\trp$.

Let us also further introduce the notation $N_w=\frac{n_w}{2}$ and
$N_y=\frac{n_y}{2}$, \[ M=\frac{1}{2}\left[
\begin{array}{cc} 1 & i
\\ 1 & -i\end{array}\right],\] and $\Gamma=P_{N_w}\diag_{N_w}(M)$.
Moreover, let $^*$ denote the adjoint of a Hilbert space operator
(by this we mean that the operator is a map from one Hilbert space
to another), and let $X^\#$ denote the operation of taking the
adjoint of each element of $X$, where $X$ is a matrix/array of
Hilbert space operators. Also, let $X^\dagger=(X^\#)\trp$.

Then we have the following definition of an open  quantum harmonic
oscillator by generalizing slightly  the linear model given in
\cite[Section 4]{EB05}:

\begin{definition}
The system (\ref{linear-c}) (with $\beta=0$) is said to be an {\em open quantum
harmonic oscillator} if $\Theta$ is canonical and there exist a
quadratic Hamiltonian $H=x(0)\trp R x(0)$, with a  real and
symmetric Hamiltonian matrix $R$ of dimension $n \times n$, and a
coupling operator $L=\Lambda x(0)$, with complex-valued coupling
matrix $\Lambda$ of dimension $n_w \times n$, such that:
\begin{eqnarray*}
x_k(t)= U(t)^*x_k(0)U(t), \quad k=1,\ldots,n, \quad y_l(t) =
U(t)^*w_l(t)U(t), \quad l=1,\ldots,n_y,
\end{eqnarray*}
where $\{U(t);t \geq 0\}$ is an adapted process of unitary
operators satisfying the following QSDE \cite[Section 2.5]{EB05} :
\begin{eqnarray*}
dU(t)&{=}&(-iH dt-\frac{1}{2}L^\dagger L dt+[\smhsp -L^\dagger
\smhsp L\trp\smhsp]\Gamma dw(t))U(t), \smhsp U(0)=I.
\end{eqnarray*}
In this case the matrices $A,B,C,D$ are given by:
\begin{eqnarray}
A&=&2\Theta(R+\Im(\Lambda^\dagger \Lambda)); \label{eq_coef_cond_A}\\
B&=&2i\Theta[\smhsp -\Lambda^\dagger \smhsp \Lambda\trp
\smhsp]\Gamma;
\label{eq_coef_cond_B} \\
C&=&P_{N_y}\trp \left[\begin{array}{cc} \Sigma_{N_y} &
0_{N_y \times N_w} \\ 0_{N_y \times N_w} & \Sigma_{N_y}  \end{array}\right]\left[\begin{array}{c} \Lambda+\Lambda^\# \\
-i\Lambda+i\Lambda^\# \end{array} \right]; \label{eq_coef_cond_C} \\
D&=&P_{N_y}\trp \left[\begin{array}{cc} \Sigma_{N_y} & 0_{N_y \times
N_w}
\\ 0_{N_y \times N_w} & \Sigma_{N_y}
\end{array}\right]P_{N_w}=[\begin{array}{cc} I_{n_y \times n_y} &
0_{n_y \times (n_w-n_y)} \end{array}] \label{eq_coef_cond_D},
\end{eqnarray}
where $\Sigma_{N_y}=[\smhsp I_{N_y \times N_y} \smhsp
0_{N_y \times (N_w-N_y)}\smhsp]$.
\end{definition}

\subsection{Augmentation of a linear QSDE}
\label{sec:qsde_aug}

If $\Theta$ is degenerate canonical then we
may perform an augmentation in which $\Theta$ is embedded into a
larger skew symmetric matrix $\tilde \Theta$ which is canonical
\emph{up to permutation} (this means $\tilde \Theta$ becomes
canonical after permutation of appropriate rows and columns). To
do this, let
$\theta=[\Theta_{ij}]_{i,j=n'+1,\ldots,n}=\diag_{\frac{n-n'}{2}}(J)$
if $n'<n$. Here $\diag_{m}(J)$ denotes a $m \times m$ block
diagonal matrix with $m$ matrices $J$ on the diagonal. Define:
\[
\tilde \Theta= \left[
\begin{array}{ccc} 0_{n' \times n'} & 0_{n'
\times (n-n')} & I_{n' \times n'} \\
0_{(n-n') \times n'}  & \theta & 0_{(n-n') \times n'} \\
 - I_{n' \times n'} & 0_{n' \times (n-n')} & 0_{n'
\times n'}
\end{array} \right],
\]
where the middle block of rows is dropped whenever $n=n'$. Then by
definition $\tilde \Theta$ is canonical up to permutation and
contains $\Theta$ as a sub-matrix by removing appropriate rows and
columns of $\tilde{\Theta}$. Let $\tilde n=n+n'$, the dimension of
the rows and columns of $\tilde \Theta$. 
Define the vector $\tilde
x(t)=[x_1(t)\smhsp x_2(t) \smhsp \ldots \smhsp x_{n}(t) \smhsp
z_1(t) \smhsp z_2(t) \smhsp \ldots \smhsp z_{n'}(t) ]\trp$ of
variables. We now define the
following linear QSDE
\begin{eqnarray}
\label{eq_aug_qsde}
\begin{array}{rl}
d\tilde x(t) &=  \left[\begin{array}{cc} A & 0_{n \times n_c} \\
A' &
A'' \end{array} \right]\tilde x(t)dt+\left[\begin{array}{c} B \\
B'\end{array} \right]dw(t),   \\
\tilde y(t) &= \left[\begin{array}{cc} C & C' \end{array}\right]
\tilde x(t) dt + Ddw(t) \end{array}
\end{eqnarray} where
 $A'$, $A''$, $B'$ and $C'$ are, respectively, some real $n'
\times n$, $n' \times n' $, $n' \times n_w$ and $n_y \times n'$
matrices, and the initial variables $\tilde x(0)$ satisfy the commutation relations
$ \tilde x_0 \tilde
x(0)\trp-( \tilde x(0) \tilde x(0) \trp)\trp=2i\tilde \Theta$.
We shall refer to the system (\ref{eq_aug_qsde}) as an
\emph{augmentation} of (\ref{linear-c}).

\begin{remark}
\label{rm_phys_real_aug} In the proof of Theorem
\ref{thm_phys_real} it is shown that the augmentation can be
chosen to preserve commutation relations whenever the original
system does.
\end{remark}


\subsection{Formal definition of physical realizability}
\label{sec:formal-phys-real} With open quantum harmonic
oscillators and augmentations having been defined, we are now
ready to introduce a formal definition of physical realizability
of the QSDE (\ref{linear-c}). A discussion regarding the
definition follows after Theorem \ref{thm_phys_real} in which
necessary and sufficient conditions for physical realizability are
given.

\begin{definition}
\label{df_phys_real} The system (\ref{linear-c}) is said to be
physically realizable if one of the following holds:

\begin{enumerate}

\item $\Theta$ is   canonical  and (\ref{linear-c})
 represents the dynamics of an open quantum harmonic
oscillator.

\item $\Theta$ is degenerate canonical and there exists an
augmentation (\ref{eq_aug_qsde}) which, after a suitable
relabelling of the components $\tilde x_1(t),\ldots,\tilde
x_{\tilde n}(t)$ of $\tilde x(t)$, represents the dynamics of an
open quantum harmonic oscillator.
\end{enumerate}

\end{definition}

The following theorem, whose proof is given in the appendix, provides
 necessary and sufficient conditions
for physical realizability.

\begin{theorem}  \label{thm_phys_real}
The system (\ref{linear-c}) is physically
realizable if and only if:
\begin{eqnarray}
iA\Theta + i\Theta A \trp + BT_wB\trp&=&0
\label{eq_realize_cond_A}, \\
B\left[\begin{array}{c} I_{n_y \times n_y} \\
0_{(n_w-n_y) \times n_y}
\end{array} \right]&=&\Theta C\trp P_{N_y}
\trp\left[\begin{array}{cc} 0_{N_y \times N_y} & I_{N_y \times N_y} \\ -I_{N_y
\times N_y} & 0_{N_y \times N_y} \end{array} \right]P_{N_y}=\Theta
C\trp\diag_{N_y}(J) \label{eq_realize_cond_B},
\end{eqnarray}
and $D$ satisfies (\ref{eq_coef_cond_D}). Moreover for canonical
$\Theta$, the Hamiltonian and coupling matrices have explicit
expressions as follows. The Hamiltonian matrix $R$ is uniquely
given by $R=\frac{1}{4}(-\Theta A+A\trp \Theta)$, and the coupling
matrix $\Lambda$ is given uniquely by
\begin{eqnarray}
\Lambda=-\frac{1}{2}i\left[\begin{array}{cc} 0_{N_w \times N_w} &
I_{N_w \times N_w}\end{array} \right] (\Gamma^{-1})\trp B\trp
\Theta.
\end{eqnarray}
In the case that $\Theta$ is degenerate canonical, a physically
realizable augmentation of the system can be constructed  to
determine the associated Hamiltonian and coupling operators using
the above explicit formulas.
\end{theorem}

\begin{remark}
Note that the Hamiltonian and coupling operators are determined by (\ref{eq_realize_cond_A}), while conditions (\ref{eq_coef_cond_D}) and (\ref{eq_realize_cond_B}) relate to the required form of the output equation.
\end{remark}

\begin{remark}  \label{rmk:efficient}
It is possible to consider  the problem of realization more broadly than discussed above by including additional components, such as beam splitters and phase shifts that   commonly occur in quantum optics. While Theorem \ref{thm_phys_real} characterizes the existence of physically realizable controllers, detailed development of an efficient realization methodology is beyond the scope of the present paper.
\end{remark}

\section{Dissipation Properties}
\label{sec:perform-diss}

In this section, we describe various dissipation properties  frequently
used in control engineering, suitably adapted to the quantum
context. These properties  concern the influence of disturbance inputs
on energy transfers and stability. In particular, we give a quantum version of
the Strict Bounded Real Lemma (Corollary \ref{C3}) which will be
employed in section \ref{sec:synth-H} for quantum $H^\infty$ controller
synthesis. In this section, we consider the following quantum system
of the form (\ref{linear-c}):
\begin{eqnarray}
dx(t) &=& Ax(t) dt + [\begin{array}{cc}B &
G\end{array}][\begin{array}{cc} dw(t)\trp & dv(t)\trp
\end{array}]\trp;
\nonumber \\
dz(t) &=& C x(t) dt + [\begin{array}{cc}D &
H\end{array}][\begin{array}{cc} dw(t)\trp & dv(t)\trp
\end{array}]\trp\label{linear-diss}
\end{eqnarray}
In this quantum system, the input channel has two components, $dw =
\beta_w dt + d\tilde w$ which represents   disturbance signals, and
$dv$, which represents   additional noise sources.

\begin{definition}
\label{D1}
Given an operator valued quadratic form
\[
r(x,\beta_w) = [x^T  \beta_w^T ]R\left[\begin{array}{l} x \\\beta_w\end{array}\right]
\]
where
\[
R =\left[\begin{array}{ll} R_{11} & R_{12} \\ R_{12}^T  & R_{22}
\end{array}\right]
\]
 is a given real symmetric matrix,
we say the system (\ref{linear-diss}) is {\em dissipative with supply rate}
$r(x,\beta_w)$ if there exists a positive operator valued quadratic form
$V(x) = x^T X x$ (where $X$ is a real positive definite symmetric
matrix) and a constant
$\lambda >  0$ such that
\begin{eqnarray}
\langle V(x(t))\rangle + \int_0^t\langle
  r(x(s),\beta_w(s))\rangle ds
 \leq  \langle V(x(0))\rangle + \lambda t ~~~\forall t >  0,   \label{diss}
\end{eqnarray}
for all Gaussian states $\rho$ for the initial variables $x(0)$.
Here we use the shorthand notation $\langle \cdot \rangle$ for expectation over all initial variables and noises, for both the plant and the controller.

We say that the system (\ref{linear-diss}) is {\em strictly
dissipative} if there exists a constant $\epsilon > 0$ such that
inequality (\ref{diss}) holds with the matrix $R$ replaced by the
matrix $R + \epsilon I$.
\end{definition}

The term $\langle V(x(t)) \rangle$ serves as a generalization to
quantum stochastic systems (\ref{linear-diss}) of the notion of
an abstract internal energy for the system at time $t$. On the
other hand, the term $\langle r(x(t),\beta_w(t))\rangle$ is a
quantum generalization of the notion of abstract power flow
into and out of the system at time $t$. Both of these are notions
which are widely used in the stability analysis of linear and
non-linear deterministic systems \cite{WIL72,VDS96}. The
dissipation inequality (\ref{diss}) is a generalization of the
corresponding inequality that was introduced for classical 
stochastic systems  in  \cite{DJP00}, see
\cite{DJ06}.

The following theorem, whose proof is given in the appendix,
relates the property of dissipativeness to certain linear matrix
inequalities.

\begin{theorem}
\label{T1}
Given a quadratic form $r(x,\beta_w)$ defined as above, then the quantum
stochastic system (\ref{linear-diss}) is dissipative with supply rate
$r(x,\beta_w)$
if and only if there exists a real positive definite symmetric matrix
$X$ such that the following matrix inequality
is satisfied:
\begin{equation}
\left(\begin{array}{cc} { A}^T X+X{ A}+R_{11} & R_{12}+X{ B}\\{ B}^T X+R_{12}^T  &R_{22}\end{array}\right) \leq 0 .
\label{MI-1}
\end{equation}
Furthermore, the system is strictly dissipative if and only if
there exists a real positive definite symmetric matrix $X$ such
that the following matrix inequality is satisfied:
\begin{equation}
\left(\begin{array}{cc} { A}^T X+X{ A}+R_{11} & R_{12}+X{ B}\\{ B}^T X+R_{12}^T  &R_{22}\end{array}\right) < 0 .
\label{MI-2}
\end{equation}

Moreover, if either of (\ref{MI-1}) or (\ref{MI-2}) holds then the
required constant $\lambda \geq  0$ can be chosen as $\lambda=\lambda_0$, where
\begin{equation}
\label{equation-lambda}
\lambda_0 =\tr\left[\left[\begin{array}{l} { B}^T\\
      { G}^T\end{array}\right]X \left[\begin{array}{ll} { B} &
       { G}\end{array}\right]F\right] .
\end{equation}
Here, the matrix $F$ is defined by the following relation:
\begin{equation}
\label{ito-F}
Fdt = \left[\begin{array}{l}d w \\ d  v
  \end{array}\right] \left[\begin{array}{ll} d w^T & d   v^T
  \end{array}\right].
\end{equation}
\end{theorem}

We now present some corollaries to the above theorem corresponding to
special cases of the matrix $R$ defined in terms of the error output
operator $\beta_z(t)=Cx(t) + D\beta_w(t)$.
\begin{definition}
\label{D2}
The quantum stochastic system (\ref{linear-diss}) is said to
be {\em Bounded Real with disturbance attenuation $g$} if the system (\ref{linear-diss}) is dissipative with
supply rate
\begin{eqnarray*}
r(x,\beta_w) &=& \beta_z^T \beta_z - g^2 \beta_w^T \beta_w
= [x^T  \beta_w^T ]\left[\begin{array}{cc} C^T C & C^T { D} \\ { D}^T C &
    { D}^T D -g^2I
\end{array}\right]\left[\begin{array}{l} x\\\beta_w\end{array}\right].
\end{eqnarray*}
Also, the quantum stochastic system (\ref{linear-diss}) is said to
be {\em Strictly Bounded Real with disturbance attenuation $g$} if the system (\ref{linear-diss}) is strictly
dissipative with this supply rate.
\end{definition}

Using the  above definition of a bounded real system, we obtain
the following corollary from Theorem \ref{T1}. (e.g., see also
\cite{BEFB94} for the corresponding classical result.)

\begin{corollary}
\label{C2}
The quantum stochastic system (\ref{linear-diss}) is bounded
real with disturbance attenuation $g$ if and only if
there exists a positive definite symmetric matrix $X \in \rbb^{n\times
  n}$ such that the following matrix inequality is
satisfied:
\[
\left(\begin{array}{cc} { A}^T X+X{ A}+{ C}^T { C} & { C}^T { D}+X{ B}\\{ B}^T X+{ D}^T { C}
    &{ D}^T { D}-g^2I\end{array}\right) \leq 0.
\]
Furthermore, the quantum stochastic system  is strictly
bounded real  with disturbance attenuation $g$ if
and only if there
exists a positive definite symmetric matrix $X \in \rbb^{n\times
  n}$ such that the following matrix inequality is satisfied:
\[
\left(\begin{array}{cc} { A}^T X+X{ A}+{ C}^T { C} & { C}^T { D}+X{ B}\\{ B}^T X+{ D}^T { C}
    &{ D}^T { D}-g^2I\end{array}\right) <  0.
\]
Moreover, in both cases the required constant $\lambda \geq  0$ can be
chosen as $\lambda=\lambda_0$, where $\lambda_0$ is defined by (\ref{equation-lambda}).
 \end{corollary}

Now combining this corollary with the standard Strict Bounded Real
Lemma (e.g., see \cite{PAJ91,ZK88}) we obtain the following Corollary.

\begin{corollary}
\label{C3}
The following statements are equivalent
\begin{enumerate}[(i)]
\item
The quantum stochastic system (\ref{linear-diss}) is strictly
bounded real with disturbance attenuation $g$.
\item
${ A}$ is a stable matrix and $\|{ C}(sI-{ A})^{-1}{ B}+{ D}\|_\infty <  g.$
\footnote{The $H^\infty$ norm notation used here is standard \cite{ZDG96}, and applies to the classical transfer function $C(sI-{ A})^{-1}{ B}+{ D}$, {\em not} the quantum system (\ref{linear-diss}). In this paper we do not define nor use transfer functions for quantum systems.}
\item
$g^2I-{ D}^T { D} >  0$ and there exists a positive definite matrix $
\tilde X > 0$ such that
\begin{eqnarray*}
{ A}^T  \tilde X+ \tilde X{ A}+{ C}^T { C}
+ ( \tilde X{ B}+{ C}^T { D})(g^2I-{ D}^T { D})^{-1}({ B}^T  \tilde X+{ D}^T { C}) <  0.
\end{eqnarray*}
\item
$g^2I-{ D}^T { D} >  0$ and the algebraic Riccati equation
\begin{eqnarray*}
{ A}^T  X+ X{ A}+{ C}^T { C}
+ ( X{ B}+{ C}^T { D})(g^2I-{ D}^T { D})^{-1}({ B}^T  X+{ D}^T { C})
= 0
\end{eqnarray*}
has a stabilizing solution $X \geq 0$.
\end{enumerate}
Furthermore, if these statements hold then $X <   \tilde X$.
\end{corollary}

\section{$H^\infty$ Controller Synthesis}
\label{sec:synth-H}

In this section, we consider the problem of $H^\infty$ controller
design for quantum systems. As we shall see, we do
not restrict ourselves to classical controllers.
  The closed loop plant-controller system  is defined in Subsection
  \ref{sec:models-closed}, and then in Subsection
  \ref{sec:synth-H-main} we apply the Strict Bounded Real Lemma to the
  closed loop system to obtain our main results.  In Subsection,
  \ref{sec:synth-H-real} we provide conditions under which a
  controller is physically realizable.

\subsection{The Closed Loop Plant-Controller System}
\label{sec:models-closed}

The general linear model \er{linear-c} described above is the
prototype for the interconnection of components which will make up the
quantum control system. In control system design, we prescribe a
system called the
{\em plant}, and seek to find another system, called a {\em
  controller}, in such a way that desired closed loop behavior is
achieved. We now introduce our plant and controller models,
and the resulting closed loop.

We consider {\em plants} described by non-commutative stochastic models
of the following form defined in an analogous way to the quantum
system (\ref{linear-c}):
 \begin{eqnarray}
\label{sys} dx(t) &=& Ax(t)dt + [\begin{array}{ccc} B_0 & B_1 &
 B_2 \end{array}][\begin{array}{ccc} dv(t)\trp & dw(t)\trp &
du(t)\trp\end{array}]\trp;
\quad x(0)=x_0;
\nonumber \\
dz(t) &=& C_1x(t)dt + D_{12}du(t); \nonumber \\
dy(t) &=& C_2x(t)dt +[\begin{array}{ccc} D_{20} & D_{21} & 0_{n_y
\times n_u}
\end{array}][\begin{array}{ccc} dv(t)\trp & dw(t)\trp & du(t)\trp
\end{array}]\trp.
\end{eqnarray}
Here $x(t)$ is a vector of plant variables. The input $w(t)$ is
represents a  {\em disturbance} signal of the form
\er{w-general}. The signal $u(t)$ is a {\em control} input of the
form
\begin{equation}
du(t) = \beta_u(t)dt + d\tilde u(t)
\label{u-quant}
\end{equation}
where $\tilde u(t)$ is the noise part of $u(t)$ and $\beta_u(t)$
is the adapted, self-adjoint finite variation part of $u(t)$.
Also, $dv(t)$ represents any additional quantum noise in the
plant. The vectors $v(t)$, $\tilde w(t)$ and $\tilde u(t)$ are
quantum noises with Ito matrices $F_v$, $F_{\tilde w}$ and
$F_{\tilde u}$ which are all non-negative Hermitian.

{\em Controllers} are assumed to be non-commutative stochastic systems of the form
\begin{eqnarray}
\label{controller}
d\xi(t) &=& A_K\xi(t)dt +
[\begin{array}{cc} B_{K1} &B_K \end{array}][\begin{array}{cc} dv_K(t)\trp & dy(t)\trp\end{array}]\trp; \quad \xi(0)=\xi_0 \nonumber \\
du(t) &=& C_K\xi(t)dt + [\begin{array}{cc} B_{K0} & 0_{n_u \times
n_y}\end{array}][\begin{array}{cc} dv_K(t)\trp &
dy(t)\trp\end{array}]\trp
\end{eqnarray}
where $ \xi(t) = [\begin{array}{ccc} \xi_1(t) & \ldots &
\xi_{n_K}(t)\end{array}]\trp $ is a vector of self-adjoint
controller variables. The noise \\ $ v_K(t) = [\begin{array}{ccc}
{v_K}_1(t) & \ldots & {v_K}_{K_v}(t)
\end{array}]\trp $ is a vector of non-commutative Wiener
processes (in vacuum states) with non-zero Ito products as in
\er{w-Ito-general} and with canonical Hermitian Ito matrix
$F_{v_K}$. 

At time $t=0$, we also assume that $x(0)$ commutes with $\xi(0)$.
The {\em closed loop system} is obtained by making the
identification $\beta_u(t)=C_K\xi(t)$ and interconnecting \er{sys}
and \er{controller} to give
\begin{eqnarray}
\label{cl1}
d\eta(t) &=& \left[\begin{array}{ll}A & B_2C_K \\B_K C_2 & A_K
  \end{array}\right]\eta(t) dt
+ \left[\begin{array}{ll}B_0 & B_2B_{K0} \\B_KD_{20} &
    B_{K1}\end{array}\right]
\left[\begin{array}{l}
dv(t) \\dv_{K}(t)
\end{array}
\right]
+\left[\begin{array}{l} B_1\\B_K D_{21} \end{array}\right]dw(t);
\nonumber \\
dz(t) &=& \left[\begin{array}{ll}C_1 &
    D_{12}C_K\end{array}\right]\eta(t) dt
+ \left[\begin{array}{ll}0 & D_{12}B_{K0} \end{array}\right]
\left[\begin{array}{l}
dv(t) \\dv_{K}(t)
\end{array}
\right]
\end{eqnarray}
where $ \eta(t) = [\begin{array}{cc}x(t)\trp & \xi(t)\trp
\end{array}]\trp$.  That is, we can write
\begin{eqnarray}
\label{cl2}
d\eta(t) &=& \tilde A \eta(t) dt + \tilde Bdw(t) + \tilde Gd\tilde
v(t) = \tilde A \eta(t) dt
+ \left[\begin{array}{ll}\tilde B & \tilde G
  \end{array}\right]\left[\begin{array}{l}
dw(t)\\ d\tilde v(t)\end{array}\right]; \nonumber  \\
dz(t) &=& \tilde C\eta(t) dt + \tilde H d\tilde v(t)
= \tilde C\eta(t) dt + \left[\begin{array}{ll}0 & \tilde
    H\end{array}\right]
\left[\begin{array}{l}
dw(t)\\ d\tilde v(t)\end{array}\right]
\end{eqnarray}
where
\begin{eqnarray*}
\tilde v(t) &=& \left[\begin{array}{l}
v(t) \\v_{K}(t)
\end{array}
\right];~~\tilde A = \left[\begin{array}{ll}A & B_2C_K \\B_K C_2 & A_K
  \end{array}\right];
\tilde B = \left[\begin{array}{l} B_1\\B_K
    D_{21} \end{array}\right];~~\tilde G = \left[\begin{array}{ll}B_0 & B_2B_{K0} \\B_KD_{20} &
    B_{K1}\end{array}\right]; \nonumber \\
\tilde C &=& \left[\begin{array}{ll}C_1 &
    D_{12}C_K\end{array}\right];~~ \tilde H = \left[\begin{array}{ll}0
    & D_{12}B_{K0} \end{array}\right].
\end{eqnarray*}
Note that the closed loop system (\ref{cl2}) is a system of the
form (\ref{linear-c}). 

\subsection{$H^{\infty}$ control objective}

The goal of the $H^{\infty}$ controller synthesis is to find a
controller (\ref{controller}) such that for a \emph{given}
disturbance attenuation parameter $g>0$:
\begin{equation}
\int_{0}^{t} \langle z(s)\trp z(s)  + \epsilon x(s)\trp x(s) \rangle ds \leq  (g^2-\epsilon^2)  \int_{0}^{t}
\langle \beta_w(s)\trp \beta_w(s) \rangle ds + \mu_1+\mu_2 t,
\label{h-infty-objective}
\end{equation}
is satisfied for some real constants $\epsilon, \mu_1,\mu_2 >0$. Thus  the controller bounds the effect of the
``energy'' of the signal $\beta_{w}(t)$ and the noise variances on
the ``energy'' of the signal $z(t)$. 

Necessary and sufficient conditions for the existence
of a specific type of controller which achieves this goal for a
given $g$ are given in the next section, as well as explicit
formulas for $A_K$, $B_K$ and $C_K$. The results parallel the
corresponding well-known results for classical linear systems
(see, e.g., \cite{DGKF89,PAJ91}).


\subsection{Necessary and Sufficient Conditions}
\label{sec:synth-H-main}

In order to present our  results on quantum $H^\infty$ control, we
will require that the plant system (\ref{sys}) satisfies the following
assumptions.
\begin{assumption}
\label{A1}
\
\begin{enumerate}
\item
$D_{12}^TD_{12} = E_1 > 0.$
\item
$D_{21}D_{21}^T = E_2 > 0. $
\item
The matrix $\left[\begin{array}{ll} A - j\omega I & B_2 \\
C_1 & D_{12}\end{array}\right]$ is full rank for all $\omega \geq 0$.
\item
The matrix $\left[\begin{array}{ll} A - j\omega I & B_1 \\
C_2 & D_{21}\end{array}\right]$ is full rank for all $\omega \geq 0$.
\end{enumerate}
\end{assumption}

Our   results will be stated in terms of the following pair of
algebraic Riccati equations:
\begin{eqnarray}
\label{r1}
(A-B_2E_1^{-1}D_{12}^TC_1)^TX
+X(A-B_2E_1^{-1}D_{12}^TC_1)
+X(B_1B_1^T - g^2B_2E_1^{-1}B_2')X
\nonumber \\
+g^{-2}C_1^T(I-D_{12}E_1^{-1}D_{12}^T)C_1 = 0;
\end{eqnarray}

\begin{eqnarray}
\label{r2}
(A-B_1D_{21}^TE_2^{-1}C_2)Y+Y(A-B_1D_{21}^TE_2^{-1}C_2)
+Y(g^{-2}C_1^TC_1-C_2^TE_2^{-1}C_2)Y
\nonumber \\
+ B_1(I-D_{21}^TE_2^{-1}D_{21})B_1^T =
0.
\end{eqnarray}

The solutions to these Riccati equations will be required to satisfy
the following assumption.
\begin{assumption}
\label{A2}
\
\begin{enumerate}[(i)]
\item
$
A-B_2E_1^{-1}D_{12}^TC_1 +(B_1B_1^T - g^2B_2E_1^{-1}B_2')X $ is a
stability matrix.
\item
$A-B_1D_{21}^TE_2^{-1}C_2+Y(g^{-2}C_1^TC_1-C_2^TE_2^{-1}C_2)$ is a stability
matrix.
\item
The matrix $XY$ has a spectral radius strictly less than one.
\end{enumerate}
\end{assumption}

Our results will show that if the Riccati equations (\ref{r1}),
(\ref{r2}) have solutions satisfying Assumption \ref{A2}, then a
controller of the form (\ref{controller}) will solve the $H^\infty$
control problem under consideration if its system matrices are
constructed from the
Riccati solutions as follows:
\begin{eqnarray}
\label{controller_matrices}
A_K &=& A+B_2C_K-B_KC_2+(B_1-B_KD_{21})B_1^TX; \nonumber \\
B_K &=& (I-YX)^{-1}(YC_2^T+B_1D_{21}^T)E_2^{-1}; \nonumber \\
C_K &=& -E_1^{-1}(g^{2}B_2^TX+D_{12}^TC_1).
\end{eqnarray}

We are now in a position to present our main result concerning
$H^\infty$ controller synthesis.  

\begin{theorem}
\label{thm:H-synth}
{\em Necessity.} Consider the system (\ref{sys}) and suppose that
Assumption \ref{A1} is satisfied. If there exists a controller of the
form (\ref{controller}) such that the resulting closed loop system
(\ref{cl2}) is
strictly bounded real with disturbance attenuation $g$, then the
Riccati equations (\ref{r1}),
(\ref{r2}) will have stabilizing solutions $X \geq 0$ and $Y \geq 0$
satisfying Assumption \ref{A2}.

{\em Sufficiency.} Suppose the Riccati equations (\ref{r1}),
(\ref{r2}) have stabilizing solutions $X \geq 0$ and $Y \geq 0$
satisfying Assumption \ref{A2}. If the controller (\ref{controller})
is such that the matrices $A_K$, $B_K$, $C_K$ are as defined in
(\ref{controller_matrices}), then the resulting closed loop system
(\ref{cl2}) will be strictly bounded real with disturbance attenuation
$g$. Also the constant $\lambda
\geq 0$ in Definition \ref{D1} can be chosen as in
(\ref{equation-lambda}), $\lambda=\lambda_0$,  where the matrix $\tilde P$ is
as defined in Lemma \ref{L1} for the closed loop system.
\end{theorem}

The controller parameters $B_{K0}$, $B_{K1}$, and the controller
noise $v_K$ are not given in the construction described in the
sufficiency part of Theorem \ref{thm:H-synth}. They are free as
far as the $H^\infty$ objective is concerned. In the next
subsection, we show how they may be chosen to give a controller
that is physically realizable.

\subsection{Physical realization of controllers}
\label{sec:synth-H-real}

We now show  that given an {\em arbitrary} choice of commutation matrix $\Theta_K$ for the controller,
it is always possible to find a physically realizable controller  in the sense of
Definition \ref{df_phys_real}. This means that the controller can be
chosen to be purely quantum, purely classical, or a combination of
quantum and classical components. 

\begin{theorem}
\label{thm_realization_hin}
Assume $F_y = D_{20}F_v D_{20}^T + D_{21}F_w D_{21}^T$ is canonical.
 Let $\{A_K,B_K,C_K\}$ be an arbitrary triple (such as given by (\ref{controller_matrices})), and select
 the  controller commutation matrix $\Theta_K$ to be canonical or degenerate canonical, as desired. Then there exists  controller parameters $B_{K0}$, $B_{K1}$, and the controller noise $v_K$ such that the controller
 (\ref{controller}) is physically realizable. In particular, $2 i \Theta_K= \xi(t)\xi(t)\trp-(\xi(t)\xi(t)\trp)\trp$
 for all $t \geq 0$ whenever $2 i \Theta_K= \xi(0)\xi(0)\trp-(\xi(0)\xi(0)\trp)\trp$.
 \end{theorem}

The proof of this theorem  depends on the following lemma for the case
in which $\Theta_K$ is canonical. For the degenerate canonical case, this
lemma can be applied to an augmentation of the controller.
We shall use the notation of
Section \ref{sec:qho}, and as in the  discussion in Section \ref{sec:models},  we
may take $B_K$ to
have an even number of columns and $C_K$ to have an even number of
rows.

\begin{lemma}
\label{lemma_realization_hin} Let $F_y$ be canonical and $\{A_K,B_K,C_K\}$ be  such that
$A_K \in \rbb^{n_K \times n_K}$, $B_K \in \rbb^{n_K \times
m_K}$, $C_K \in \rbb^{l_k \times n_K}$, $n_K=2N_{\xi}$, $m_K
=2N_{y}$ and $l_K =2N_u$ for positive integers $N_{\xi}$, $N_y$
and $N_u$, and  $\Theta_K=\diag_{N_\xi}(J)$ is canonical.
 Then there exists
an integer $N_{v_K} \geq N_u$ and $B_{K1} \in \rbb^{n_K \times
2N_{w_K}}$, with $N_{w_K} = N_{v_K}+N_y$, such that the system (\ref{controller})
is physically realizable with
\[
B_{K0}=P_{N_u}\trp \left[ \begin{array}{cc} \Sigma_{N_u} & 0_{N_u
\times  N_{w_K}} \\ 0_{N_u \times N_{w_K}} & \Sigma_{N_u}
\end{array}\right]P_{N_{w_K}}
\left[\begin{array}{c} I_{2N_{v_K} \times 2N_{v_K}} \\ 0_{m_k
\times 2N_{v_K}}\end{array} \right]=[\begin{array}{cc}I_{n_y
\times n_y} & 0_{n_y \times (n_w-n_y)}\end{array}],
\]
\begin{eqnarray}
R&=&\frac{1}{2}(Z+Z\trp); \label{eq_R} \\
B_{K1}&=& \left[\begin{array}{cc} B_{K1,1} & B_{K1,2}
\end{array}\right]; \label{eq_BK1} \\
\Lambda&=&\left[\begin{array}{ccc} \frac{1}{2}C_K \trp
P_{N_u}\trp\left[\begin{array}{c} I \\ iI
\end{array}\right]& \Lambda_{b1}\trp &
\Lambda_{b2}\trp
\end{array}\right]\trp; \label{eq_Lambda} \\
B_{K1,1}&=&-i\Theta_K C_K\trp \diag_{N_u}(iJ); \label{eq_def_BK11}\\
\Lambda_{b2}&=&-i\left[\begin{array}{cc} I_{N_y \times N_y} &
0_{N_y \times N_y}\end{array}\right]P_{N_y}\diag_{N_y}(M) B_K\trp \Theta_K; \label{eq_b2_BK}\\
B_{K1,2}&=& 2i\Theta_K \left[\begin{array}{cc} -\Lambda_{b1}^\dagger
& \Lambda_{b1}\trp \end{array}
\right]P_{N_{v_K}-N_u}\diag_{N_{v_K}-N_u}(M) \label{eq_def_BK12}
\end{eqnarray}
where $Z=-\frac{1}{2}\Theta A_K$ and  $N_{v_K} \geq N_u+1$.  Here
$\Lambda_{b1}$ is any complex $(N_{v_K}-N_u) \times n_K$ matrix
such that
\begin{eqnarray}
\Lambda_{b1}^\dagger \Lambda_{b1} &=&\Xi +
i\left(\frac{1}{2}(Z-Z\trp)-\frac{1}{4}C_K\trp P_{N_u}\trp \left[
\begin{array}{cc}  0 & I
\\ -I & 0 \end{array} \right]P_{N_u} C_K-\Im
(\Lambda_{b2}^\dagger \Lambda_{b2})\right) \label{eq_Lambda_b1_2}
\end{eqnarray}
where $\Xi$ is any real symmetric $n_K \times n_K$ matrix such
that the right hand side of (\ref{eq_Lambda_b1_2}) is non-negative
definite.
\end{lemma}

The proof of Lemma \ref{lemma_realization_hin} is given in the appendix.

\begin{remark} Note that the condition $N_{v_K} \geq N_u$ is
significant since it implies that there is no direct feedthrough of the
signal $y(t)$ to $u(t)$ as required for (\ref{controller}).
For compatibility between the equations (\ref{controller}) and
(\ref{sys}), it is necessary that the corresponding Ito matrices
satisfy the following condition:
\begin{eqnarray}
F_u=B_{K0}F_{v_K}B_{K0}\trp. \label{ccr-controller-2}
\end{eqnarray}
However, since $F_{v_K}$ and $F_{u}$ are, by convention, in
canonical form, (\ref{ccr-controller-2}) is always satisfied. To
see this, we simply note that the $2N_u$ elements of $B_{K0} v_K$
are a subset of pairs of conjugate real and imaginary quadratures
in $v_K$. Hence it follows that if $F_{v_K}$ is canonical then
$F_u$ must also be canonical and (\ref{ccr-controller-2}) is
 automatically satisfied.
\end{remark}

\section{Robust Stability}
\label{sec:appeg-robust}

The $H^\infty$ control approach of Section \ref{sec:synth-H} leads to
a closed loop quantum system of the form (\ref{cl2}) which is strictly
bounded real with disturbance attenuation $g$. We now show that this
property can be used to guarantee stability
robustness against real parameter uncertainties. Indeed, we will suppose
that the true closed loop quantum system corresponding to the system
(\ref{cl2}) is described by the equations
\begin{equation}
\label{cl2_Delta}
d\eta(t) = \bar A \eta(t)dt + \tilde
Gd\tilde v(t)
\end{equation}
where $\bar A = \tilde A+ \tilde B \Delta  \tilde C$ and
$\Delta$ is a constant but unknown uncertainty matrix satisfying
\begin{equation}
\label{Delta_bound}
\Delta^{T} \Delta \leq \frac{1}{g^2}I.
\end{equation}

\begin{definition}
The closed loop quantum system (\ref{cl2_Delta}) is said to be {\em
  mean square stable} if there exists a real positive definite matrix
$X > 0$ and a constant $\lambda > 0$ such that
\[
 \langle \eta(t)^TX\eta(t)\rangle + \int_0^t \langle
\eta(s)^T\eta(s)\rangle ds \leq \langle \eta(0)^TX\eta(0)
\rangle + \lambda t
\quad \forall t >0
\]
for all Gaussian states $\rho$.
\end{definition}

The following lemma and theorem relates the robust stability of the
above system to its $H^\infty$ properties. The proofs of this lemma and
this theorem can be found in the appendix.

\begin{lemma}
\label{lem_MSS}
The quantum system (\ref{cl2_Delta}) is mean square stable if and only
if the matrix $\bar A$ is a stable matrix.
\end{lemma}

\begin{theorem}
\label{small_gain}
If the closed loop quantum system (\ref{cl2}) is strictly
bounded real with disturbance attenuation $g$, then the true closed
loop system (\ref{cl2_Delta}) is
mean square stable for all $\Delta$ satisfying (\ref{Delta_bound}).
\end{theorem}

\section{$H^\infty$ Synthesis in Quantum Optics}
\label{sec:appeg}

Quantum optics is an important area in quantum physics and quantum technology
and provides a  promising means of implementing quantum information
and computing devices; e.g., see \cite{KLM01}. In this section we give some
examples of controller design for   simple quantum optics plants based on optical cavities and optical amplifiers coupled to   optical fields; e.g., see
\cite{BR04,GZ00}.  We give explicit realizations of controllers which are fully quantum, fully classical, and mixed quantum-classical using standard quantum optical components and electronics.

\subsection{Quantum Controller Synthesis}
\label{sec:qoptics-qc}

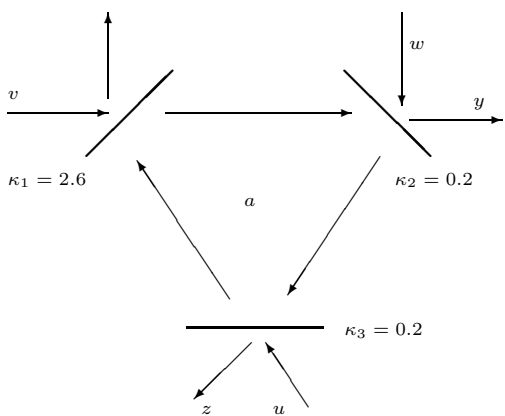
\begin{figure}[h]
\begin{center}

\setlength{\unitlength}{2368sp}%
\begingroup\makeatletter\ifx\SetFigFont\undefined%
\gdef\SetFigFont#1#2#3#4#5{%
  \reset@font\fontsize{#1}{#2pt}%
  \fontfamily{#3}\fontseries{#4}\fontshape{#5}%
  \selectfont}%
\fi\endgroup%
\begin{picture}(5199,4261)(1564,-4910)
\put(5101,-4036){\makebox(0,0)[lb]{\smash{{\SetFigFont{7}{8.4}{\familydefault}{\mddefault}{\updefault}{\color[rgb]{0,0,0}$\kappa_3=0.2$}%
}}}}
\thicklines
{\color[rgb]{0,0,0}\put(5101,-1261){\line( 1,-1){900}}
}%
\thinlines
{\color[rgb]{0,0,0}\put(3226,-1711){\vector( 1, 0){1950}}
}%
{\color[rgb]{0,0,0}\put(5476,-2161){\vector(-2,-3){969.231}}
}%
{\color[rgb]{0,0,0}\put(3901,-3661){\vector(-2, 3){969.231}}
}%
\thicklines
{\color[rgb]{0,0,0}\put(3451,-3961){\line( 1, 0){1425}}
}%
\thinlines
{\color[rgb]{0,0,0}\put(1576,-1711){\vector( 1, 0){1050}}
}%
{\color[rgb]{0,0,0}\put(2626,-1561){\vector( 0, 1){900}}
}%
{\color[rgb]{0,0,0}\put(4126,-4111){\vector(-1,-1){600}}
}%
{\color[rgb]{0,0,0}\put(4726,-4786){\vector(-2, 3){450}}
}%
{\color[rgb]{0,0,0}\put(5776,-1786){\vector( 1, 0){975}}
}%
{\color[rgb]{0,0,0}\put(5701,-661){\vector( 0,-1){975}}
}%
\put(1576,-1561){\makebox(0,0)[lb]{\smash{{\SetFigFont{7}{8.4}{\familydefault}{\mddefault}{\updefault}{\color[rgb]{0,0,0}$v$}%
}}}}
\put(5776,-1036){\makebox(0,0)[lb]{\smash{{\SetFigFont{7}{8.4}{\familydefault}{\mddefault}{\updefault}{\color[rgb]{0,0,0}$w$}%
}}}}
\put(4051,-2686){\makebox(0,0)[lb]{\smash{{\SetFigFont{7}{8.4}{\familydefault}{\mddefault}{\updefault}{\color[rgb]{0,0,0}$a$}%
}}}}
\put(6451,-1636){\makebox(0,0)[lb]{\smash{{\SetFigFont{7}{8.4}{\familydefault}{\mddefault}{\updefault}{\color[rgb]{0,0,0}$y$}%
}}}}
\put(4351,-4861){\makebox(0,0)[lb]{\smash{{\SetFigFont{7}{8.4}{\familydefault}{\mddefault}{\updefault}{\color[rgb]{0,0,0}$u$}%
}}}}
\put(3601,-4861){\makebox(0,0)[lb]{\smash{{\SetFigFont{7}{8.4}{\familydefault}{\mddefault}{\updefault}{\color[rgb]{0,0,0}$z$}%
}}}}
\put(1576,-2461){\makebox(0,0)[lb]{\smash{{\SetFigFont{7}{8.4}{\familydefault}{\mddefault}{\updefault}{\color[rgb]{0,0,0}$\kappa_1=2.6$}%
}}}}
\put(5626,-2461){\makebox(0,0)[lb]{\smash{{\SetFigFont{7}{8.4}{\familydefault}{\mddefault}{\updefault}{\color[rgb]{0,0,0}$\kappa_2=0.2$}%
}}}}
\thicklines
{\color[rgb]{0,0,0}\put(2401,-2161){\line( 1, 1){900}}
}%
\end{picture}%

\caption{An optical cavity (plant).}
\label{fig:cavity1}
\end{center}
\end{figure}

We consider an optical cavity resonantly coupled to three optical
channels $v$, $w$, $u$ as in  Figure \ref{fig:cavity1}. The  control objective is to attenuate the effect of the disturbance signal $w$ on the output $z$---physically this means to dim the light emerging from $z$ resulting from light shone in at $w$.

The dynamics of this cavity system is described by the evolution of
its annihilation operator $a$   (representing a
traveling wave).
In the quadrature notation of  (\ref{sys}),
$x_1(t)=q(t)=a(t)+a^\ast(t)$,  $x_2(t)=p(t)=(a(t)-a^\ast(t)/i$,
$v(t) = (v_1(t) ,v_2(t))^T$, $w(t) = (w_1(t) ,w_2(t))^T$, $u(t)
= (u_1(t) ,  u_2(t))^T$.  The quantum noises $v$, $\tilde w$ have
Hermitian Ito
matrices $F_v=F_{\tilde w}=I+iJ$. This leads to a system of the form (\ref{sys}) with the
following system matrices:
\begin{eqnarray}
\label{system_matrices} A &=& -\frac{\gamma}{2} I; ~~ B_0 =
-\sqrt{\kappa_1} I;~~B_1= -\sqrt{\kappa_2} I;~~B_2
= -\sqrt{\kappa_3} I;\nonumber \\
&& \quad (\gamma=\kappa_1+\kappa_2+\kappa_3) \nonumber \\
C_1 &=& \sqrt{\kappa_3} I;~~D_{12} = I; \nonumber \\
C_2 &=& \sqrt{\kappa_2} I; ~~ D_{21} = I.
\end{eqnarray}
In this model, the boson commutation relation
$[a,a^\ast]=1$ holds. This means that the commutation matrix for this plant is $\Theta_P=J$.

In our example, we will choose the total cavity decay rate $\kappa
= 3$ and the coupling coefficients $\kappa_1 =2.6$,
$\kappa_2=\kappa_3=0.2$. With a disturbance attenuation constant
of $g=0.1$, it was found that the Riccati equations (\ref{r1}) and
(\ref{r2}) have stabilizing solutions satisfying Assumption
\ref{A2}. These Riccati solutions were as follows: $X=Y=0_{2\times
2}$. Then, it follows from Theorem \ref{T1} that  if a controller
of the form (\ref{controller}) is applied to this system with
matrices $A_K$, $B_K$, $C_K$ defined as in
(\ref{controller_matrices}) then the resulting closed loop system
will be strictly bounded real with disturbance attenuation $g$. In
our case, these matrices are given by
$$A_K=-1.1 I, ~ B_K=-0.447 I, ~ C_K=-0.447 I .
$$

\begin{figure}[h]
\begin{center}

\setlength{\unitlength}{2368sp}%
\begingroup\makeatletter\ifx\SetFigFont\undefined%
\gdef\SetFigFont#1#2#3#4#5{%
  \reset@font\fontsize{#1}{#2pt}%
  \fontfamily{#3}\fontseries{#4}\fontshape{#5}%
  \selectfont}%
\fi\endgroup%
\begin{picture}(5262,5620)(3376,-10619)
\put(7426,-8311){\makebox(0,0)[lb]{\smash{{\SetFigFont{7}{8.4}{\familydefault}{\mddefault}{\updefault}{\color[rgb]{0,0,0}$\kappa_{K3}=0.2$}%
}}}}
\thinlines
{\color[rgb]{0,0,0}\put(3451,-7561){\vector( 1, 0){1050}}
}%
\thicklines
{\color[rgb]{0,0,0}\put(6938,-7148){\line( 1,-1){900}}
}%
{\color[rgb]{0,0,0}\put(5401,-9736){\line( 1, 0){1425}}
}%
\thinlines
{\color[rgb]{0,0,0}\put(5101,-7711){\vector( 1, 0){1950}}
}%
{\color[rgb]{0,0,0}\put(7254,-7923){\vector(-2,-3){969.231}}
}%
{\color[rgb]{0,0,0}\put(5915,-9443){\vector(-2, 3){969.231}}
}%
{\color[rgb]{0,0,0}\put(7651,-7636){\vector( 1, 0){975}}
}%
{\color[rgb]{0,0,0}\put(5926,-9886){\vector(-1,-1){600}}
}%
{\color[rgb]{0,0,0}\put(6786,-10538){\vector(-2, 3){450}}
}%
{\color[rgb]{0,0,0}\put(4426,-6586){\framebox(600,750){}}
}%
{\color[rgb]{0,0,0}\put(4726,-7411){\vector( 0, 1){825}}
}%
{\color[rgb]{0,0,0}\put(4726,-7036){\makebox(2.7778,19.4444){\SetFigFont{5}{6}{\rmdefault}{\mddefault}{\updefault}.}}
}%
{\color[rgb]{0,0,0}\put(4726,-6811){\makebox(2.7778,19.4444){\SetFigFont{5}{6}{\rmdefault}{\mddefault}{\updefault}.}}
}%
{\color[rgb]{0,0,0}\put(4726,-5836){\vector( 0, 1){825}}
}%
{\color[rgb]{0,0,0}\put(7351,-5011){\vector( 0,-1){2325}}
}%
\put(6001,-8536){\makebox(0,0)[lb]{\smash{{\SetFigFont{7}{8.4}{\familydefault}{\mddefault}{\updefault}{\color[rgb]{0,0,0}$a_K$}%
}}}}
\put(6901,-10561){\makebox(0,0)[lb]{\smash{{\SetFigFont{7}{8.4}{\familydefault}{\mddefault}{\updefault}{\color[rgb]{0,0,0}$v_{K2}$}%
}}}}
\put(4501,-6286){\makebox(0,0)[lb]{\smash{{\SetFigFont{7}{8.4}{\rmdefault}{\mddefault}{\updefault}{\color[rgb]{0,0,0}Phase}%
}}}}
\put(4501,-6511){\makebox(0,0)[lb]{\smash{{\SetFigFont{7}{8.4}{\rmdefault}{\mddefault}{\updefault}{\color[rgb]{0,0,0}Shift}%
}}}}
\put(4501,-6061){\makebox(0,0)[lb]{\smash{{\SetFigFont{7}{8.4}{\rmdefault}{\mddefault}{\updefault}{\color[rgb]{0,0,0}$180^\circ$}%
}}}}
\put(4351,-5536){\makebox(0,0)[lb]{\smash{{\SetFigFont{7}{8.4}{\familydefault}{\mddefault}{\updefault}{\color[rgb]{0,0,0}$u$}%
}}}}
\put(6901,-5536){\makebox(0,0)[lb]{\smash{{\SetFigFont{7}{8.4}{\familydefault}{\mddefault}{\updefault}{\color[rgb]{0,0,0}$y$}%
}}}}
\put(3451,-7411){\makebox(0,0)[lb]{\smash{{\SetFigFont{7}{8.4}{\familydefault}{\mddefault}{\updefault}{\color[rgb]{0,0,0}$v_{K1}$}%
}}}}
\put(3376,-8311){\makebox(0,0)[lb]{\smash{{\SetFigFont{7}{8.4}{\familydefault}{\mddefault}{\updefault}{\color[rgb]{0,0,0}$\kappa_{K1}=0.2$}%
}}}}
\put(3751,-9811){\makebox(0,0)[lb]{\smash{{\SetFigFont{7}{8.4}{\familydefault}{\mddefault}{\updefault}{\color[rgb]{0,0,0}$\kappa_{K2}=1.8$}%
}}}}
\thicklines
{\color[rgb]{0,0,0}\put(4313,-8048){\line( 1, 1){900}}
}%
\end{picture}%

\caption{An optical cavity  quantum realization of the controller ($\Theta_K=J$) for the plant shown in Figure \ref{fig:cavity1}.}
\label{fig:cavity2}
\end{center}
\end{figure}
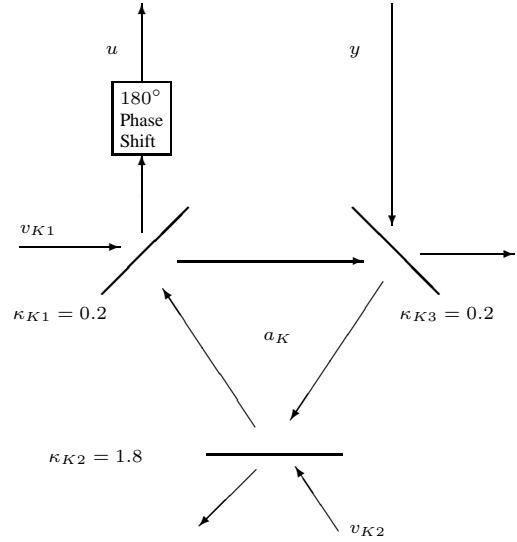

In this case, the controller (\ref{controller}) can be implemented
with another optical cavity with annihilation operator $a_K$ (with
quadratures $\xi_1=q_K=a_K+a_K^\ast$,
$\xi_2=p_K=(a_K-a_K^\ast)/i$, $\xi=(q_K,p_K)^T$), corresponding to
$\Theta_K=J$, connected at the output with a $180^{\rm o}$ phase
shifter (see Remark \ref{rmk:efficient}). The controller cavity
has coupling coefficients $\kappa_{K1}=0.2$, $\kappa_{K2}=1.8$,
$\kappa_{K3}=0.2$, and $\kappa_{K}=2.2$ and is a physically
realizable system with dynamics:
\begin{eqnarray*}
d\xi(t)&=&A_K \xi(t)dt + [\begin{array}{cc} B_{K1} & B_K \end{array}][\begin{array}{cc} dv_K\trp & dy\trp \end{array}]\trp\\
d\tilde u(t)&=& -C_K \xi(t) dt + [\begin{array}{cc} I_{2 \times 2}
& 0_{2 \times 4} \end{array}][\begin{array}{cc} dv_K\trp & dy\trp
\end{array}]\trp,
\end{eqnarray*}
where $B_{K1} =[\begin{array}{cc} -0.447 I & -1.342 I$$
\end{array}]$,
$v_K(t)=(v_{K11}(t) ,v_{K12}(t)  , v_{K21}(t)  ,v_{K22}(t)
y(t))^T$ are the quadratures of two independent canonical quantum
noise sources, and $\tilde u(t)$ is the output of the cavity. The
overall output of the controller is $u(t)$, given by $u(t)=K_s
\tilde u(t)$, where $K_s=-I_{2 \times 2}$. Here $K_s$ models the
$180^{\rm o}$ phase shift at the output of the cavity. Thus, the
overall controller (an optical cavity cascaded with a $180^{\rm
o}$ phase shifter) is of the form (\ref{controller}) with $B_{K0}
= [ -\begin{array}{cc} I & 0    \end{array}]$ and $B_{K1}$ as
given before. This controller is illustrated in Figure
\ref{fig:cavity2}.


\subsection{Robust Stability in Quantum Optics}
\label{sec:qoptics-robust}
We now modify the above example to allow for uncertainty in one of the
optical cavity parameters using the results of Section
\ref{sec:appeg-robust}. Indeed, we consider the same set up as in
Figure \ref{fig:cavity1} and  assume that there is uncertainty in the
value of the coupling coefficient $\kappa_1$ corresponding to the optical channel
$v$. In this case, the equations  (\ref{sys}) describing the
optical cavity now have matrices
\begin{eqnarray}
\label{system_matrices_delta}
A &=& -\frac{\gamma+\delta}{2} I; ~~ B_0 = \sqrt{\kappa_1+\delta} I;~~B_1=
-\sqrt{\kappa_2} I;~~B_2
= -\sqrt{\kappa_3} I;\nonumber \\
C_1 &=& \sqrt{\kappa_3} I;~~D_{12} = I; \nonumber \\
C_2 &=& \sqrt{\kappa_2} I; ~~ D_{21} = I.
\end{eqnarray}
This is our true system which depends on the unknown parameter
$\delta$.

In order to apply our $H^\infty$ theory together with the results of Section
\ref{sec:appeg-robust} to this system, we must overbound the
uncertainty in the matrix $A$. Indeed, let $S$ be any non-singular
matrix.  If $|\delta| \leq \mu$, then we can write
$
-\frac{\delta}{2} I = \tilde B_1 \Delta \tilde C_1
$
where $\tilde B_1 = \frac{\mu}{2} S$, $\tilde C_1 = S^{-1}$ and
$\Delta = -\frac{\delta}{\mu}I$ satisfies $\Delta^T\Delta \leq
I$. Hence, if we consider a family of systems of the form (\ref{sys}) with the
 system matrices
\begin{eqnarray}
\label{system_matrices_Delta} A &=& -\frac{\gamma}{2} I + \tilde
B_1 \Delta \tilde C_1; ~~ B_0 = \sqrt{\kappa_1+\delta} I;~~B_1=
-\sqrt{\kappa_2} I;~~B_2
= -\sqrt{\kappa_3} I;\nonumber \\
C_1 &=& \sqrt{\kappa_3} I;~~D_{12} = I; \nonumber \\
C_2 &=& \sqrt{\kappa_2} I; ~~ D_{21} = I
\end{eqnarray}
where $\Delta^T\Delta \leq I$, this will include the true system. Now,
in order to apply the result of Section
\ref{sec:appeg-robust} to this problem, we consider the $H^\infty$
problem defined by a system of the form (\ref{sys}) where
\begin{eqnarray}
\label{system_matrices_robust_Hinf}
A &=& -\frac{\gamma}{2} I; ~~ B_0 = \sqrt{\kappa_1+\delta} I;~~B_{10}=
-\sqrt{\kappa_2} I;~~B_{1} =\left[\begin{array}{ll}B_{10} & \tilde
    B_1\end{array}\right]; \nonumber \\
B_2
&=& -\sqrt{\kappa_3} I;\nonumber \\
C_{10} &=& \sqrt{\kappa_3} I;~~C_1 = \left[\begin{array}{l} C_{10} \\ g
    \tilde C_1\end{array}\right];~~D_{120} = I;~~D_{12}=
\left[\begin{array}{c} D_{120} \\ 0\end{array}\right];  \nonumber \\
C_2 &=& \sqrt{\kappa_2} I; ~~ D_{210} =
I;~~D_{21}=\left[\begin{array}{cc} D_{210} & 0 \end{array}\right].
\end{eqnarray}
Here $g$ is the disturbance attenuation parameter in the $H^\infty$
control problem to be considered. Note that the
matrix $B_0$ depends on the unknown parameter $\delta$. However, this
matrix is not involved in the calculation of the $H^\infty$
controller.

As in the original example, we will choose the nominal cavity decay rate $\gamma =
3$ and the nominal coupling coefficients of $\kappa_1 =2.6$, $\kappa_2
= \kappa_3 = 0.2$. Also, we let
$\mu = 0.1$. That is, we are considering a $10\%$ variation in the
coupling coefficient. With a disturbance
attenuation constant of $g=0.1$ and $S =
1.5 I$, it was found that the Riccati
equations (\ref{r1}) and (\ref{r2}) have stabilizing solutions
satisfying Assumption \ref{A2}. These Riccati solutions were as
follows: $X=0.1733 I$, $Y=0.0022 I$.
Also, the corresponding controller
matrices were given by
\begin{eqnarray}
\label{example_matrices_u}
A_K = -1.0997I,  ~
B_K = -0.4464 I, ~
C_K =-0.4464 I  .
\end{eqnarray}

Now as in the original example, the controller defined by the matrices
(\ref{example_matrices_u}) can be implemented by another optical
cavity.
In this case,  $\kappa_{K1}=0.1993$, $\kappa_{K2}=1.8008$,
$\kappa_{K3}=0.1993$, and $\gamma_{K}=2.1993$. As in the original
example, the controller is illustrated in Figure \ref{fig:cavity2}.

It follows from Theorem \ref{T1} that  the
resulting closed loop system  satisfies the strict
bounded real condition with disturbance attenuation $g$. Indeed, this
closed loop system will be described by the equations (\ref{cl2})
where
\begin{eqnarray*}
 \tilde A &=& \left[\begin{array}{ll}A & B_2C_K \\B_K C_2 & A_K
  \end{array}\right]; \nonumber \\
\tilde B &=& \left[\begin{array}{l} B_1\\B_K
    D_{21} \end{array}\right] = \left[\begin{array}{ll}
B_{10} & \tilde B_1 \\
B_KD_{210} & 0\end{array}\right];~~
\tilde G = \left[\begin{array}{ll}B_0 & B_2B_{K0} \\B_KD_{20} &
    B_{K1}\end{array}\right]; \nonumber \\
\tilde C &=& \left[\begin{array}{ll}C_1 &
    D_{12}C_K\end{array}\right]
= \left[\begin{array}{ll}
C_{10} &  D_{120}C_K\\ g\tilde C_1 & 0
\end{array}\right];~~ \tilde D = \left[\begin{array}{ll}0
    & D_{12}B_{K0} \end{array}\right].
\end{eqnarray*}
Now, since this system is strictly bounded real with disturbance
attenuation $g$, it follows from
Corollary \ref{C3} that $\|\tilde C (sI-\tilde A)^{-1}\tilde B\|_\infty <
g$. From this, we can conclude that
\begin{equation}
\label{nominal_attenuation}
\left\| \left[\begin{array}{ll}
C_{10} &  D_{120}C_K
\end{array}\right](sI-\tilde A)^{-1}\left[\begin{array}{l}
B_{10} \\
B_KD_{210} \end{array}\right]\right\|_\infty <
g
\end{equation}
and
\begin{equation}
\label{robustness_SBR}
\left\| \left[\begin{array}{ll}
\tilde C_1 & 0
\end{array}\right](sI-\tilde A)^{-1}\left[\begin{array}{l}
\tilde B_1 \\
 0 \end{array}\right]\right\|_\infty <1.
\end{equation}
Using Corollary \ref{C3}, (\ref{nominal_attenuation}) implies that the
nominal closed loop system  strictly bounded real
with disturbance attenuation $g$. Also, (\ref{robustness_SBR}) implies that the closed loop
system
\begin{eqnarray*}
d\eta(t) &=& \left[\begin{array}{ll}A & B_2C_K \\B_K C_2 & A_K
  \end{array}\right]\eta(t) dt
+ \left[\begin{array}{ll}B_0 & B_2B_{K0} \\B_KD_{20} &
    B_{K1}\end{array}\right]
\left[\begin{array}{l}
dv(t) \\dv_{K}(t)
\end{array}
\right]
+\left[\begin{array}{l} \tilde B_1\\0 \end{array}\right]dw(t);
\nonumber \\
dz(t) &=& \left[\begin{array}{ll}\tilde C_1 &
    0\end{array}\right]\eta(t) dt
\end{eqnarray*}
is strictly bounded real with unity disturbance attenuation. From this, it follows from Theorem
\ref{small_gain} that the
closed loop uncertain system
\begin{eqnarray*}
d\eta(t) &=& \left[\begin{array}{ll}A+\tilde B_1 \Delta \tilde C_1 & B_2C_K \\B_K C_2 & A_K
  \end{array}\right]\eta(t) dt
+ \left[\begin{array}{ll}B_0 & B_2B_{K0} \\B_KD_{20} &
    B_{K1}\end{array}\right]
\left[\begin{array}{l}
dv(t) \\dv_{K}(t)
\end{array}
\right]
\end{eqnarray*}
is mean square stable for all matrices $\Delta$ such that $\Delta^T
\Delta \leq I$. Hence, we can conclude that the true closed loop
system is mean square stable.

Note that for this example, it is also possible to verify that the
true closed loop system must not only be mean square stable but must
also be strictly bounded real with disturbance attenuation $g$.

\subsection{Classical Controller Synthesis}
\label{sec:qoptics-cc}

 In subsection \ref{sec:qoptics-qc} we obtained a quantum controller corresponding to the choice $\Theta_K=J$. We now show that if we instead choose $\Theta_K=0$, the controller that is realized is classical, with appropriate transitions to and from the quantum plant.

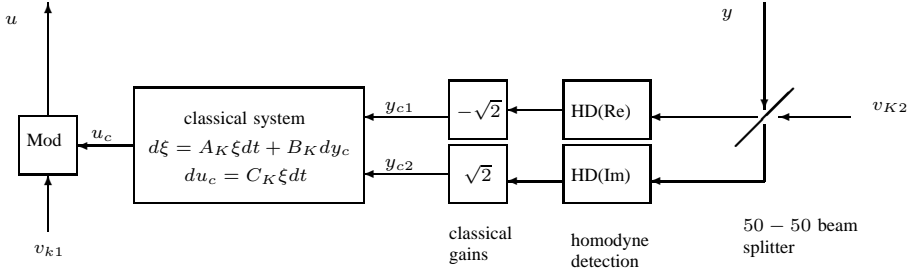
\begin{figure}[h]
\begin{center}
\setlength{\unitlength}{2368sp}%
\begingroup\makeatletter\ifx\SetFigFont\undefined%
\gdef\SetFigFont#1#2#3#4#5{%
  \reset@font\fontsize{#1}{#2pt}%
  \fontfamily{#3}\fontseries{#4}\fontshape{#5}%
  \selectfont}%
\fi\endgroup%
\begin{picture}(9859,2787)(2551,-5536)
\put(7351,-4636){\makebox(0,0)[lb]{\smash{{\SetFigFont{7}{8.4}{\familydefault}{\mddefault}{\updefault}{\color[rgb]{0,0,0}$\sqrt{2}$}%
}}}}
\thinlines
{\color[rgb]{0,0,0}\put(2701,-4561){\framebox(600,600){}}
}%
{\color[rgb]{0,0,0}\put(8401,-4636){\vector(-1, 0){600}}
}%
{\color[rgb]{0,0,0}\put(7201,-4561){\vector(-1, 0){900}}
}%
{\color[rgb]{0,0,0}\put(3901,-4261){\vector(-1, 0){600}}
}%
{\color[rgb]{0,0,0}\put(3001,-3961){\vector( 0, 1){1200}}
}%
{\color[rgb]{0,0,0}\put(3001,-5161){\vector( 0, 1){600}}
}%
{\color[rgb]{0,0,0}\put(11401,-3961){\vector(-1, 0){750}}
}%
{\color[rgb]{0,0,0}\put(10501,-2761){\vector( 0,-1){1125}}
}%
{\color[rgb]{0,0,0}\put(7201,-4861){\framebox(600,600){}}
}%
{\color[rgb]{0,0,0}\put(7201,-4186){\framebox(600,600){}}
}%
{\color[rgb]{0,0,0}\put(8401,-3886){\vector(-1, 0){600}}
}%
{\color[rgb]{0,0,0}\put(7201,-3961){\vector(-1, 0){900}}
}%
{\color[rgb]{0,0,0}\put(8401,-4186){\framebox(900,600){}}
}%
{\color[rgb]{0,0,0}\put(8401,-4861){\framebox(900,600){}}
}%
{\color[rgb]{0,0,0}\put(10426,-3961){\vector(-1, 0){1125}}
}%
{\color[rgb]{0,0,0}\put(10501,-4036){\line( 0,-1){600}}
\put(10501,-4636){\vector(-1, 0){1200}}
}%
\thicklines
{\color[rgb]{0,0,0}\put(10201,-4261){\line( 1, 1){600}}
}%
\put(4426,-4036){\makebox(0,0)[lb]{\smash{{\SetFigFont{7}{8.4}{\familydefault}{\mddefault}{\updefault}{\color[rgb]{0,0,0}classical system}%
}}}}
\put(4426,-4636){\makebox(0,0)[lb]{\smash{{\SetFigFont{7}{8.4}{\familydefault}{\mddefault}{\updefault}{\color[rgb]{0,0,0}$du_c=C_K \xi dt$}%
}}}}
\put(2851,-5386){\makebox(0,0)[lb]{\smash{{\SetFigFont{7}{8.4}{\familydefault}{\mddefault}{\updefault}{\color[rgb]{0,0,0}$v_{k1}$}%
}}}}
\put(11626,-3886){\makebox(0,0)[lb]{\smash{{\SetFigFont{7}{8.4}{\familydefault}{\mddefault}{\updefault}{\color[rgb]{0,0,0}$v_{K2}$}%
}}}}
\put(2776,-4261){\makebox(0,0)[lb]{\smash{{\SetFigFont{7}{8.4}{\familydefault}{\mddefault}{\updefault}{\color[rgb]{0,0,0}Mod}%
}}}}
\put(8476,-5311){\makebox(0,0)[lb]{\smash{{\SetFigFont{7}{8.4}{\familydefault}{\mddefault}{\updefault}{\color[rgb]{0,0,0}homodyne}%
}}}}
\put(8476,-5536){\makebox(0,0)[lb]{\smash{{\SetFigFont{7}{8.4}{\familydefault}{\mddefault}{\updefault}{\color[rgb]{0,0,0}detection}%
}}}}
\put(7201,-5236){\makebox(0,0)[lb]{\smash{{\SetFigFont{7}{8.4}{\familydefault}{\mddefault}{\updefault}{\color[rgb]{0,0,0}classical}%
}}}}
\put(7201,-5461){\makebox(0,0)[lb]{\smash{{\SetFigFont{7}{8.4}{\familydefault}{\mddefault}{\updefault}{\color[rgb]{0,0,0}gains}%
}}}}
\put(10276,-5161){\makebox(0,0)[lb]{\smash{{\SetFigFont{7}{8.4}{\familydefault}{\mddefault}{\updefault}{\color[rgb]{0,0,0}$50-50$ beam}%
}}}}
\put(10276,-5386){\makebox(0,0)[lb]{\smash{{\SetFigFont{7}{8.4}{\familydefault}{\mddefault}{\updefault}{\color[rgb]{0,0,0}splitter}%
}}}}
\put(2551,-2986){\makebox(0,0)[lb]{\smash{{\SetFigFont{7}{8.4}{\familydefault}{\mddefault}{\updefault}{\color[rgb]{0,0,0}$u$}%
}}}}
\put(10051,-2911){\makebox(0,0)[lb]{\smash{{\SetFigFont{7}{8.4}{\familydefault}{\mddefault}{\updefault}{\color[rgb]{0,0,0}$y$}%
}}}}
\put(8476,-3961){\makebox(0,0)[lb]{\smash{{\SetFigFont{7}{8.4}{\familydefault}{\mddefault}{\updefault}{\color[rgb]{0,0,0}HD(Re)}%
}}}}
\put(8476,-4636){\makebox(0,0)[lb]{\smash{{\SetFigFont{7}{8.4}{\familydefault}{\mddefault}{\updefault}{\color[rgb]{0,0,0}HD(Im)}%
}}}}
\put(3451,-4186){\makebox(0,0)[lb]{\smash{{\SetFigFont{7}{8.4}{\familydefault}{\mddefault}{\updefault}{\color[rgb]{0,0,0}$u_{c}$}%
}}}}
\put(6526,-3886){\makebox(0,0)[lb]{\smash{{\SetFigFont{7}{8.4}{\familydefault}{\mddefault}{\updefault}{\color[rgb]{0,0,0}$y_{c1}$}%
}}}}
\put(6526,-4486){\makebox(0,0)[lb]{\smash{{\SetFigFont{7}{8.4}{\familydefault}{\mddefault}{\updefault}{\color[rgb]{0,0,0}$y_{c2}$}%
}}}}
\put(4051,-4336){\makebox(0,0)[lb]{\smash{{\SetFigFont{7}{8.4}{\familydefault}{\mddefault}{\updefault}{\color[rgb]{0,0,0}$d\xi=A_K \xi dt + B_K dy_c$}%
}}}}
\put(7276,-3961){\makebox(0,0)[lb]{\smash{{\SetFigFont{7}{8.4}{\familydefault}{\mddefault}{\updefault}{\color[rgb]{0,0,0}$-\sqrt{2}$}%
}}}}
\thinlines
{\color[rgb]{0,0,0}\put(3901,-4861){\framebox(2400,1200){}}
}%
\end{picture}%

\caption{A classical realization of the controller ($\Theta_K=0$) for the plant shown in Figure \ref{fig:cavity1}. The controller includes quantum measurement and classical modulation of  optical fields.}
\label{fig:classical-0}
\end{center}
\end{figure}

 Now, suppose we choose $v_K$ to be the
quadratures of two independent noise channels (i.e.,
$F_{v_K}=I_{4\times4}+i\diag(J,J)$). Setting $\Theta_K = 0_{2\times 2}$, Eq. (\ref{eq_realize_cond_A}) and the compatibility
requirement (\ref{ccr-controller-2}) in this context results in
the following pair of equations:
\begin{eqnarray}
B_K J B_K\trp + B_{K1}\diag(J,J)B_{K1}\trp=0 \label{eq_contr_redesign_1}\\
B_{K0}(I_{4 \times 4}+i\diag(J,J))B_{K0}\trp = I+iJ
\label{eq_contr_redesign_2}.
\end{eqnarray}
In order to find $B_{K0}$ and $B_{K1}$ solving
(\ref{eq_contr_redesign_1}) and (\ref{eq_contr_redesign_2}), we assume
the following forms for $B_{K0}$ and $B_{K1}$:
\begin{eqnarray*}
B_{K0}=\left[\begin{array}{cc} \tilde{B}_{K0} & 0_{2 \times 2}
\end{array}\right]; \quad
B_{K1}=\left[\begin{array}{cc} 0_{2 \times 2} & \tilde{B}_{K1}
\end{array}\right].
\end{eqnarray*}
Since $B_K=-0.447I$, substitution of these forms into
(\ref{eq_contr_redesign_1}) and (\ref{eq_contr_redesign_2}) gives:
\begin{eqnarray*}
\tilde{B}_{K0}(I+iJ)\tilde B_{K_0}\trp=I+iJ; \quad
0.447^2J+\tilde{B}_{K1}J \tilde{B}_{K1}\trp=0.
\end{eqnarray*}
It can be readily checked, by direct substitution, that these
equations are solved by $\tilde B_{K0}=I_{2\times 2}$ and $\tilde
B_{K1}=-0.447 \tilde I$, where $\tilde I=\left[\begin{array}{cc} 1
& 0
\\ 0 & -1
\end{array} \right]$. This completely specifies the classical realization of the controller, illustrated in Figure \ref{fig:classical-0}. The quantum signal $y$ is converted to a classical signal $y_c=(y_{c1}, y_{c2})^T=(y_1-v_{K21},v_{K22}+y_2)^T$ by imperfect continuous measurement of the real and imaginary quadratures of the optical beam, implemented in Figure \ref{fig:classical-0} by a beam splitter and two homodyne detectors \cite{BR04}. The classical signal $y_c$ is processed by a classical linear system $(A_K, B_K, C_K,0)$ to produce a classical control signal $u_c$, which then modulates (displaces) a field $v_{K1}$ to produce the optical control signal $d u=u_c dt +d v_{K1}$. This classical controller achieves exactly the same $H^\infty$ performance as the quantum controller of subsection \ref{sec:qoptics-qc}.

\begin{figure}[h]
\begin{center}
\setlength{\unitlength}{2368sp}%
\begingroup\makeatletter\ifx\SetFigFont\undefined%
\gdef\SetFigFont#1#2#3#4#5{%
  \reset@font\fontsize{#1}{#2pt}%
  \fontfamily{#3}\fontseries{#4}\fontshape{#5}%
  \selectfont}%
\fi\endgroup%
\begin{picture}(6324,4261)(1564,-4910)
\put(7126,-2761){\makebox(0,0)[lb]{\smash{{\SetFigFont{8}{9.6}{\familydefault}{\mddefault}{\updefault}{\color[rgb]{0,0,0}HD(Re)}%
}}}}
\thicklines
{\color[rgb]{0,0,0}\put(5101,-1261){\line( 1,-1){900}}
}%
\thinlines
{\color[rgb]{0,0,0}\put(3226,-1711){\vector( 1, 0){1950}}
}%
{\color[rgb]{0,0,0}\put(5476,-2161){\vector(-2,-3){969.231}}
}%
{\color[rgb]{0,0,0}\put(3901,-3661){\vector(-2, 3){969.231}}
}%
\thicklines
{\color[rgb]{0,0,0}\put(3451,-3961){\line( 1, 0){1425}}
}%
\thinlines
{\color[rgb]{0,0,0}\put(1576,-1711){\vector( 1, 0){1050}}
}%
{\color[rgb]{0,0,0}\put(2626,-1561){\vector( 0, 1){900}}
}%
{\color[rgb]{0,0,0}\put(4126,-4111){\vector(-1,-1){600}}
}%
{\color[rgb]{0,0,0}\put(4726,-4786){\vector(-2, 3){450}}
}%
{\color[rgb]{0,0,0}\put(5701,-661){\vector( 0,-1){975}}
}%
{\color[rgb]{0,0,0}\put(5926,-1786){\line( 1, 0){1425}}
}%
{\color[rgb]{0,0,0}\put(7351,-2911){\vector( 0,-1){900}}
}%
{\color[rgb]{0,0,0}\put(7351,-1786){\vector( 0,-1){675}}
}%
{\color[rgb]{0,0,0}\put(6976,-2911){\framebox(900,450){}}
}%
\put(1576,-1561){\makebox(0,0)[lb]{\smash{{\SetFigFont{7}{8.4}{\familydefault}{\mddefault}{\updefault}{\color[rgb]{0,0,0}$v$}%
}}}}
\put(5776,-1036){\makebox(0,0)[lb]{\smash{{\SetFigFont{7}{8.4}{\familydefault}{\mddefault}{\updefault}{\color[rgb]{0,0,0}$w$}%
}}}}
\put(4051,-2686){\makebox(0,0)[lb]{\smash{{\SetFigFont{7}{8.4}{\familydefault}{\mddefault}{\updefault}{\color[rgb]{0,0,0}$a$}%
}}}}
\put(4351,-4861){\makebox(0,0)[lb]{\smash{{\SetFigFont{7}{8.4}{\familydefault}{\mddefault}{\updefault}{\color[rgb]{0,0,0}$u$}%
}}}}
\put(3601,-4861){\makebox(0,0)[lb]{\smash{{\SetFigFont{7}{8.4}{\familydefault}{\mddefault}{\updefault}{\color[rgb]{0,0,0}$z$}%
}}}}
\put(7576,-3361){\makebox(0,0)[lb]{\smash{{\SetFigFont{7}{8.4}{\familydefault}{\mddefault}{\updefault}{\color[rgb]{0,0,0}$y$}%
}}}}
\thicklines
{\color[rgb]{0,0,0}\put(2401,-2161){\line( 1, 1){900}}
}%
\end{picture}%

\caption{An optical cavity  (plant) with classical output. The (real) quadrature measurement
  is achieved by homodyne photodetection (HD(Re)).}
\label{fig:cavity3}
\end{center}
\end{figure}
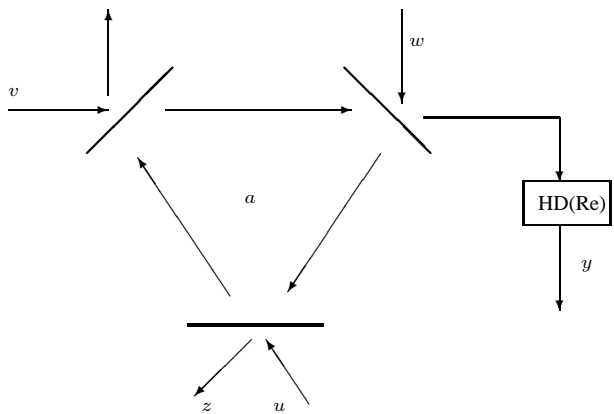

This classical controller has access to the full quantum signal $y$, and the quantum measurement occurs in the controller. The algebra based on the commutation relations enforces the quantum measurement, and also the modulation.  If we were to include measurement as part of the plant specification, then in general a different classical controller will result, with different $H^\infty$ performance. To see this, suppose that $y$ is replaced by its real quadrature in the plant specification; this situation is described by the matrices
 \begin{eqnarray}
\label{system_matrices_c}
A &=& -\frac{\gamma}{2} I; ~~ B_0 = -\sqrt{\kappa_1} I;~~B_1=
-\sqrt{\kappa_2} I;~~B_2
= -\sqrt{\kappa_3} I;\nonumber \\
C_1 &=& \sqrt{\kappa_3} I;~~D_{12} = I; \nonumber \\
C_2 &=& \sqrt{\kappa_2} \left[ \begin{array}{cc} 1 & 0  \end{array} \right]; ~~ D_{21} =  \left[ \begin{array}{cc} 1 & 0  \end{array} \right]
\end{eqnarray}
and is illustrated in Figure \ref{fig:cavity3}.
Thus the output of the plant is a classical single-variable signal.

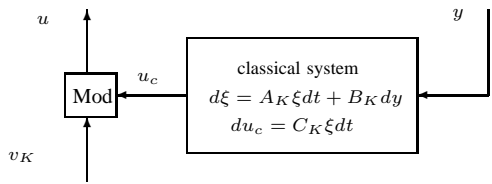
\begin{figure}[h]
\begin{center}
\setlength{\unitlength}{2368sp}%
\begingroup\makeatletter\ifx\SetFigFont\undefined%
\gdef\SetFigFont#1#2#3#4#5{%
  \reset@font\fontsize{#1}{#2pt}%
  \fontfamily{#3}\fontseries{#4}\fontshape{#5}%
  \selectfont}%
\fi\endgroup%
\begin{picture}(5037,1893)(3526,-8173)
\put(4876,-7111){\makebox(0,0)[lb]{\smash{{\SetFigFont{7}{8.4}{\familydefault}{\mddefault}{\updefault}{\color[rgb]{0,0,0}$u_c$}%
}}}}
\thinlines
{\color[rgb]{0,0,0}\put(5401,-7261){\vector(-1, 0){750}}
}%
{\color[rgb]{0,0,0}\put(4351,-7036){\vector( 0, 1){675}}
}%
{\color[rgb]{0,0,0}\put(4351,-8161){\vector( 0, 1){675}}
}%
{\color[rgb]{0,0,0}\put(5401,-7861){\framebox(2400,1200){}}
}%
{\color[rgb]{0,0,0}\put(8551,-6361){\line( 0,-1){900}}
\put(8551,-7261){\vector(-1, 0){750}}
}%
\put(4201,-7336){\makebox(0,0)[lb]{\smash{{\SetFigFont{8}{9.6}{\familydefault}{\mddefault}{\updefault}{\color[rgb]{0,0,0}Mod}%
}}}}
\put(3826,-6511){\makebox(0,0)[lb]{\smash{{\SetFigFont{7}{8.4}{\familydefault}{\mddefault}{\updefault}{\color[rgb]{0,0,0}$u$}%
}}}}
\put(3526,-7936){\makebox(0,0)[lb]{\smash{{\SetFigFont{7}{8.4}{\familydefault}{\mddefault}{\updefault}{\color[rgb]{0,0,0}$v_{K}$}%
}}}}
\put(5926,-7036){\makebox(0,0)[lb]{\smash{{\SetFigFont{7}{8.4}{\familydefault}{\mddefault}{\updefault}{\color[rgb]{0,0,0}classical system}%
}}}}
\put(5626,-7336){\makebox(0,0)[lb]{\smash{{\SetFigFont{7}{8.4}{\familydefault}{\mddefault}{\updefault}{\color[rgb]{0,0,0}$d\xi=A_K \xi dt+B_K dy$}%
}}}}
\put(5851,-7636){\makebox(0,0)[lb]{\smash{{\SetFigFont{7}{8.4}{\familydefault}{\mddefault}{\updefault}{\color[rgb]{0,0,0}$du_c=C_K \xi dt$}%
}}}}
\put(8176,-6436){\makebox(0,0)[lb]{\smash{{\SetFigFont{7}{8.4}{\familydefault}{\mddefault}{\updefault}{\color[rgb]{0,0,0}$y$}%
}}}}
{\color[rgb]{0,0,0}\put(4126,-7486){\framebox(525,450){}}
}%
\end{picture}%

\caption{Classical controller ($\Theta_K=0$) for the plant of Figure \ref{fig:cavity3}.}
\label{fig:cavity3a}
\end{center}
\end{figure}

With a disturbance
attenuation constant of $g=0.134$, it was found that the Riccati
equations (\ref{r1}) and (\ref{r2}) have the following stabilizing solutions
satisfying Assumption \ref{A2}:
\[
X = \left[\begin{array}{ll}
0& 0 \\
0 & 0 \end{array}\right];~~ Y = \left[\begin{array}{ll}
0 & 0 \\
0 & 0.121  \end{array}\right].
\]
It now follows from Theorem \ref{T1} that  if a controller of the
form (\ref{controller}) is applied to this system with the
following matrices $A_K$, $B_K$, $C_K$ defined as in
(\ref{controller_matrices}), then the resulting closed loop system
will be strictly bounded real with disturbance attenuation
$g=0.667$:
\begin{eqnarray}
\label{example_matrices1}
A_K = \left[\begin{array}{ll}
-1.1 & 0 \\
0 & -1.3 \end{array}\right];~~B_K = \left[ \begin{array}{c} -0.447 \\ 0  \end{array} \right];  
C_K = \left[\begin{array}{ll}
-0.447 & 0 \\
0 & -0.447 \end{array}\right].
\end{eqnarray}
In this case, the controller (\ref{controller}),
(\ref{example_matrices1}) is a classical system which  can be
implemented using standard electronic devices. This second classical controller is illustrated in Figure
\ref{fig:cavity3a}, and is different to the previous one. Here we have chosen
$
B_{K0}= I,
\ B_{K1}= 0,
$
and the quantum noise is canonical. The control signal is $d u=u_c dt+dv_K$, a coherent optical field.

 \subsection{Classical-Quantum Controller Synthesis}
\label{sec:qoptics-cq}

\begin{figure}[h]
\begin{center}
\setlength{\unitlength}{2368sp}%
\begingroup\makeatletter\ifx\SetFigFont\undefined%
\gdef\SetFigFont#1#2#3#4#5{%
  \reset@font\fontsize{#1}{#2pt}%
  \fontfamily{#3}\fontseries{#4}\fontshape{#5}%
  \selectfont}%
\fi\endgroup%
\begin{picture}(9174,4524)(-2411,-5173)
\put(-824,-1936){\makebox(0,0)[lb]{\smash{{\SetFigFont{7}{8.4}{\familydefault}{\mddefault}{\updefault}{\color[rgb]{0,0,0}$h$}%
}}}}
\thicklines
{\color[rgb]{0,0,0}\put(5101,-1261){\line( 1,-1){900}}
}%
\thinlines
{\color[rgb]{0,0,0}\put(3226,-1711){\vector( 1, 0){1950}}
}%
{\color[rgb]{0,0,0}\put(5476,-2161){\vector(-2,-3){969.231}}
}%
{\color[rgb]{0,0,0}\put(3901,-3661){\vector(-2, 3){969.231}}
}%
\thicklines
{\color[rgb]{0,0,0}\put(3451,-3961){\line( 1, 0){1425}}
}%
\thinlines
{\color[rgb]{0,0,0}\put(1576,-1711){\vector( 1, 0){1050}}
}%
{\color[rgb]{0,0,0}\put(2626,-1561){\vector( 0, 1){900}}
}%
{\color[rgb]{0,0,0}\put(4126,-4111){\vector(-1,-1){600}}
}%
{\color[rgb]{0,0,0}\put(5776,-1786){\vector( 1, 0){975}}
}%
{\color[rgb]{0,0,0}\put(5701,-661){\vector( 0,-1){975}}
}%
{\color[rgb]{0,0,0}\put(-1499,-4261){\framebox(2100,1500){}}
}%
{\color[rgb]{0,0,0}\put(4726,-4861){\vector(-2, 3){484.615}}
}%
{\color[rgb]{0,0,0}\put(601,-3511){\line( 1, 0){900}}
\put(1501,-3511){\line( 0,-1){1650}}
\put(1501,-5161){\line( 1, 0){3225}}
\put(4726,-5161){\line( 0, 1){375}}
}%
{\color[rgb]{0,0,0}\put(-2399,-3511){\vector( 1, 0){900}}
}%
{\color[rgb]{0,0,0}\put(-449,-1711){\vector( 0,-1){1050}}
}%
\put(1576,-1561){\makebox(0,0)[lb]{\smash{{\SetFigFont{7}{8.4}{\familydefault}{\mddefault}{\updefault}{\color[rgb]{0,0,0}$v$}%
}}}}
\put(5776,-1036){\makebox(0,0)[lb]{\smash{{\SetFigFont{7}{8.4}{\familydefault}{\mddefault}{\updefault}{\color[rgb]{0,0,0}$w$}%
}}}}
\put(4051,-2686){\makebox(0,0)[lb]{\smash{{\SetFigFont{7}{8.4}{\familydefault}{\mddefault}{\updefault}{\color[rgb]{0,0,0}$a$}%
}}}}
\put(6451,-1636){\makebox(0,0)[lb]{\smash{{\SetFigFont{7}{8.4}{\familydefault}{\mddefault}{\updefault}{\color[rgb]{0,0,0}$y$}%
}}}}
\put(4351,-4861){\makebox(0,0)[lb]{\smash{{\SetFigFont{7}{8.4}{\familydefault}{\mddefault}{\updefault}{\color[rgb]{0,0,0}$u'$}%
}}}}
\put(3601,-4861){\makebox(0,0)[lb]{\smash{{\SetFigFont{7}{8.4}{\familydefault}{\mddefault}{\updefault}{\color[rgb]{0,0,0}$z$}%
}}}}
\put(1576,-2461){\makebox(0,0)[lb]{\smash{{\SetFigFont{7}{8.4}{\familydefault}{\mddefault}{\updefault}{\color[rgb]{0,0,0}$\kappa_1=2.6$}%
}}}}
\put(5626,-2461){\makebox(0,0)[lb]{\smash{{\SetFigFont{7}{8.4}{\familydefault}{\mddefault}{\updefault}{\color[rgb]{0,0,0}$\kappa_2=0.2$}%
}}}}
\put(5101,-4036){\makebox(0,0)[lb]{\smash{{\SetFigFont{7}{8.4}{\familydefault}{\mddefault}{\updefault}{\color[rgb]{0,0,0}$\kappa_3=0.2$}%
}}}}
\put(-899,-3661){\makebox(0,0)[lb]{\smash{{\SetFigFont{7}{8.4}{\familydefault}{\mddefault}{\updefault}{\color[rgb]{0,0,0}amplifier}%
}}}}
\put(-899,-3361){\makebox(0,0)[lb]{\smash{{\SetFigFont{7}{8.4}{\familydefault}{\mddefault}{\updefault}{\color[rgb]{0,0,0}optical}%
}}}}
\put(-2399,-3811){\makebox(0,0)[lb]{\smash{{\SetFigFont{7}{8.4}{\familydefault}{\mddefault}{\updefault}{\color[rgb]{0,0,0}$u$}%
}}}}
\thicklines
{\color[rgb]{0,0,0}\put(2401,-2161){\line( 1, 1){900}}
}%
\end{picture}%

\caption{An optical amplifier-cavity system (plant).}
\label{fig:cavity4}
\end{center}
\end{figure}
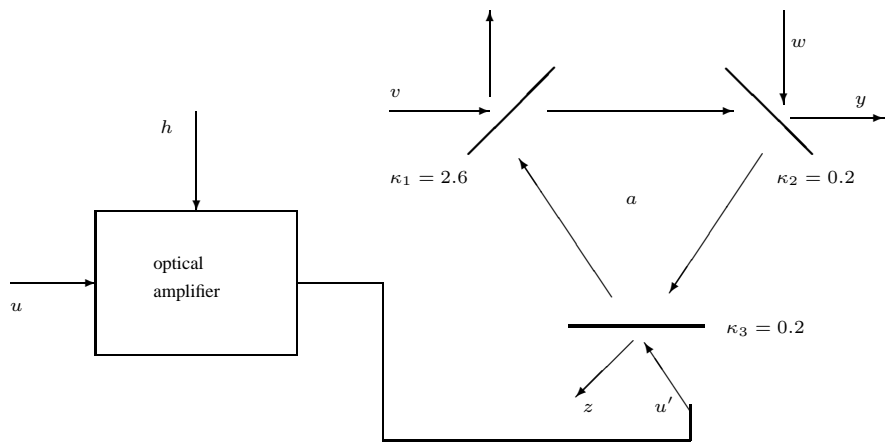

 As a final example, we illustrate the synthesis of a controller with both classical and quantum components.
 The plant has two degrees of freedom, and is formed as a cascade of
 an optical amplifier \cite{GZ00} and the cavity discussed above. This
 plant is illustrated in Figure \ref{fig:cavity4}.

The optical amplifier  has an auxiliary input $h$, which is an inverted heat bath with Ito matrix $F_h=(2N+1) I + iJ$, where $N > 0$ is a positive thermal parameter. The complete system shown in Figure \ref{fig:cavity4} is of the form (\ref{sys}) with matrices
\begin{eqnarray}
\label{system_matrices1}
A &=& \left[ \begin{array}{cc}
-\frac{\gamma}{2} I   & -\sqrt{\kappa_3\alpha}\, I
\\
0         &   -\frac{\alpha-\beta}{2} I
\end{array} \right]
; ~~ B_0 =  \left[ \begin{array}{cc}
-\sqrt{\kappa_1} I   & 0
\\
0 & \sqrt{\beta} I
\end{array} \right]
;~~B_1= \left[ \begin{array}{c}
-\sqrt{\kappa_2} I
\\
0
\end{array} \right]
;~~B_2
= \left[ \begin{array}{c}
-\sqrt{\kappa_3} I
\\
-\sqrt{\alpha} I
\end{array} \right]
;\nonumber \\
C_1 &=& \left[ \begin{array}{cc}    \sqrt{\kappa_3} I & 0 \end{array} \right]
;~~D_{12} = I; \nonumber \\
C_2 &=& \left[ \begin{array}{cc} \sqrt{\kappa_2} I & 0 \end{array} \right] ;~~ D_{20} = 0;  ~~ D_{21} = I.
\end{eqnarray}
Here $\alpha$ and $\beta$ are parameters of the optical amplifier.
The signals have Ito matrices $F_u=F_{\tilde w}=I+iJ$ and $F_v =
\mathrm{diag}( I + i J, (2N+1) I + iJ)$, and the parameters are chosen
to be $\kappa_1 =2.6$, $\kappa_2=\kappa_3=0.2$, $\alpha=1$ and $\beta=0.5$.

With a $H^\infty$ gain $g=0.1$, the Riccati
equations (\ref{r1}) and (\ref{r2}) have stabilizing solutions
satisfying Assumption \ref{A2}: $X=Y=0_{2\times 2}$. Using (\ref{controller_matrices}), the controller matrices
 $A_K$,
$B_K$, $C_K$ are
\begin{eqnarray*}
A_K = \left[\begin{array}{cc} -1.3894 I  & -0.4472 I \\
                              -0.2 I  &   -0.25 I \end{array} \right] , ~
B_K = \left[\begin{array}{c} -0.4472 I \\
                                    0_{2 \times 2} \end{array} \right] , ~
C_K = \left[\begin{array}{cc} -0.4472 I &   0_{2 \times 2}
                                     \end{array}\right] .
\end{eqnarray*}

\begin{figure}[h]
\begin{center}

 \setlength{\unitlength}{2368sp}%
\begingroup\makeatletter\ifx\SetFigFont\undefined%
\gdef\SetFigFont#1#2#3#4#5{%
  \reset@font\fontsize{#1}{#2pt}%
  \fontfamily{#3}\fontseries{#4}\fontshape{#5}%
  \selectfont}%
\fi\endgroup%
\begin{picture}(10534,6816)(-1724,-10540)
\put(3526,-4861){\makebox(0,0)[lb]{\smash{{\SetFigFont{7}{8.4}{\familydefault}{\mddefault}{\updefault}{\color[rgb]{0,0,0}$180^\circ$}%
}}}}
\thicklines
{\color[rgb]{0,0,0}\put(3601,-6961){\line( 1,-1){600}}
}%
{\color[rgb]{0,0,0}\put(4801,-7561){\line( 1, 1){600}}
}%
{\color[rgb]{0,0,0}\put(4801,-5761){\line( 1,-1){600}}
}%
\thinlines
{\color[rgb]{0,0,0}\put(4051,-6061){\vector( 1, 0){900}}
}%
{\color[rgb]{0,0,0}\put(5101,-6211){\vector( 0,-1){900}}
}%
{\color[rgb]{0,0,0}\put(4951,-7261){\vector(-1, 0){900}}
}%
{\color[rgb]{0,0,0}\put(3901,-7111){\vector( 0, 1){900}}
}%
{\color[rgb]{0,0,0}\put(3151,-9961){\framebox(3000,1200){}}
}%
{\color[rgb]{0,0,0}\put(6901,-8761){\framebox(675,600){}}
}%
{\color[rgb]{0,0,0}\put(6151,-9361){\line( 1, 0){1050}}
\put(7201,-9361){\vector( 0, 1){600}}
}%
{\color[rgb]{0,0,0}\put(7201,-8161){\line( 0, 1){900}}
\put(7201,-7261){\vector(-1, 0){2025}}
}%
{\color[rgb]{0,0,0}\put(5176,-7336){\vector( 0,-1){600}}
}%
{\color[rgb]{0,0,0}\put(3901,-7936){\vector( 0, 1){600}}
}%
{\color[rgb]{0,0,0}\put(5251,-6061){\vector( 1, 0){750}}
}%
{\color[rgb]{0,0,0}\put(3901,-5986){\vector( 0, 1){825}}
}%
{\color[rgb]{0,0,0}\put(3001,-6061){\vector( 1, 0){750}}
}%
{\color[rgb]{0,0,0}\put(8176,-8461){\vector(-1, 0){600}}
}%
{\color[rgb]{0,0,0}\put(1876,-9961){\framebox(675,525){}}
}%
{\color[rgb]{0,0,0}\put(1876,-9286){\framebox(675,525){}}
}%
{\color[rgb]{0,0,0}\put(376,-9286){\framebox(900,525){}}
}%
{\color[rgb]{0,0,0}\put(376,-9961){\framebox(900,525){}}
}%
{\color[rgb]{0,0,0}\put(3676,-7261){\line(-1, 0){4350}}
\put(-674,-7261){\vector( 0,-1){1800}}
}%
{\color[rgb]{0,0,0}\put(-599,-9061){\vector( 1, 0){975}}
}%
{\color[rgb]{0,0,0}\put(1276,-9061){\vector( 1, 0){600}}
}%
{\color[rgb]{0,0,0}\put(2551,-9061){\vector( 1, 0){600}}
}%
{\color[rgb]{0,0,0}\put(-674,-9136){\line( 0,-1){600}}
\put(-674,-9736){\vector( 1, 0){1050}}
}%
{\color[rgb]{0,0,0}\put(1276,-9736){\vector( 1, 0){600}}
}%
{\color[rgb]{0,0,0}\put(2551,-9736){\vector( 1, 0){600}}
}%
\thicklines
{\color[rgb]{0,0,0}\put(-974,-8761){\line( 1,-1){600}}
}%
\thinlines
{\color[rgb]{0,0,0}\put(-1349,-9061){\vector( 1, 0){600}}
}%
{\color[rgb]{0,0,0}\put(3451,-5161){\framebox(900,525){}}
}%
{\color[rgb]{0,0,0}\put(3901,-4561){\vector( 0, 1){825}}
}%
{\color[rgb]{0,0,0}\put(5101,-3811){\vector( 0,-1){2250}}
}%
\put(4276,-6736){\makebox(0,0)[lb]{\smash{{\SetFigFont{7}{8.4}{\familydefault}{\mddefault}{\updefault}{\color[rgb]{0,0,0}$a_K$}%
}}}}
\put(3976,-9736){\makebox(0,0)[lb]{\smash{{\SetFigFont{7}{8.4}{\familydefault}{\mddefault}{\updefault}{\color[rgb]{0,0,0}$d\zeta_c=\xi_cdt$}%
}}}}
\put(3376,-9436){\makebox(0,0)[lb]{\smash{{\SetFigFont{7}{8.4}{\familydefault}{\mddefault}{\updefault}{\color[rgb]{0,0,0}$d\xi_c=-0.25 \xi_c dt-0.1355 d\eta_c$}%
}}}}
\put(4051,-9136){\makebox(0,0)[lb]{\smash{{\SetFigFont{7}{8.4}{\familydefault}{\mddefault}{\updefault}{\color[rgb]{0,0,0}classical system}%
}}}}
\put(7051,-8536){\makebox(0,0)[lb]{\smash{{\SetFigFont{7}{8.4}{\familydefault}{\mddefault}{\updefault}{\color[rgb]{0,0,0}Mod}%
}}}}
\put(6526,-9586){\makebox(0,0)[lb]{\smash{{\SetFigFont{7}{8.4}{\familydefault}{\mddefault}{\updefault}{\color[rgb]{0,0,0}$\zeta_c$}%
}}}}
\put(526,-9061){\makebox(0,0)[lb]{\smash{{\SetFigFont{7}{8.4}{\familydefault}{\mddefault}{\updefault}{\color[rgb]{0,0,0}HD(Re)}%
}}}}
\put(526,-9736){\makebox(0,0)[lb]{\smash{{\SetFigFont{7}{8.4}{\familydefault}{\mddefault}{\updefault}{\color[rgb]{0,0,0}HD(Im)}%
}}}}
\put(1951,-9061){\makebox(0,0)[lb]{\smash{{\SetFigFont{7}{8.4}{\familydefault}{\mddefault}{\updefault}{\color[rgb]{0,0,0}$-\sqrt{2}$}%
}}}}
\put(2026,-9736){\makebox(0,0)[lb]{\smash{{\SetFigFont{7}{8.4}{\familydefault}{\mddefault}{\updefault}{\color[rgb]{0,0,0}$\sqrt{2}$}%
}}}}
\put(2776,-5911){\makebox(0,0)[lb]{\smash{{\SetFigFont{7}{8.4}{\familydefault}{\mddefault}{\updefault}{\color[rgb]{0,0,0}$v_{K1}$}%
}}}}
\put(3601,-8161){\makebox(0,0)[lb]{\smash{{\SetFigFont{7}{8.4}{\familydefault}{\mddefault}{\updefault}{\color[rgb]{0,0,0}$v_{K2}$}%
}}}}
\put(-1724,-9286){\makebox(0,0)[lb]{\smash{{\SetFigFont{7}{8.4}{\familydefault}{\mddefault}{\updefault}{\color[rgb]{0,0,0}$v_{K3}$}%
}}}}
\put(1576,-7111){\makebox(0,0)[lb]{\smash{{\SetFigFont{7}{8.4}{\familydefault}{\mddefault}{\updefault}{\color[rgb]{0,0,0}$\eta_q$}%
}}}}
\put(2701,-8836){\makebox(0,0)[lb]{\smash{{\SetFigFont{7}{8.4}{\familydefault}{\mddefault}{\updefault}{\color[rgb]{0,0,0}$\eta_{c1}$}%
}}}}
\put(2701,-9586){\makebox(0,0)[lb]{\smash{{\SetFigFont{7}{8.4}{\familydefault}{\mddefault}{\updefault}{\color[rgb]{0,0,0}$\eta_{c2}$}%
}}}}
\put(-1274,-10186){\makebox(0,0)[lb]{\smash{{\SetFigFont{7}{8.4}{\familydefault}{\mddefault}{\updefault}{\color[rgb]{0,0,0}50 - 50 beam}%
}}}}
\put(-1274,-10411){\makebox(0,0)[lb]{\smash{{\SetFigFont{7}{8.4}{\familydefault}{\mddefault}{\updefault}{\color[rgb]{0,0,0}splitter}%
}}}}
\put(451,-10261){\makebox(0,0)[lb]{\smash{{\SetFigFont{7}{8.4}{\familydefault}{\mddefault}{\updefault}{\color[rgb]{0,0,0}homodyne}%
}}}}
\put(451,-10486){\makebox(0,0)[lb]{\smash{{\SetFigFont{7}{8.4}{\familydefault}{\mddefault}{\updefault}{\color[rgb]{0,0,0}detection}%
}}}}
\put(1951,-10261){\makebox(0,0)[lb]{\smash{{\SetFigFont{7}{8.4}{\familydefault}{\mddefault}{\updefault}{\color[rgb]{0,0,0}classical}%
}}}}
\put(1951,-10486){\makebox(0,0)[lb]{\smash{{\SetFigFont{7}{8.4}{\familydefault}{\mddefault}{\updefault}{\color[rgb]{0,0,0}gains}%
}}}}
\put(5401,-6586){\makebox(0,0)[lb]{\smash{{\SetFigFont{7}{8.4}{\familydefault}{\mddefault}{\updefault}{\color[rgb]{0,0,0}$\kappa_{K4}=0.2$}%
}}}}
\put(5326,-7636){\makebox(0,0)[lb]{\smash{{\SetFigFont{7}{8.4}{\familydefault}{\mddefault}{\updefault}{\color[rgb]{0,0,0}$\kappa_{K3}=0.2$}%
}}}}
\put(2500,-7561){\makebox(0,0)[lb]{\smash{{\SetFigFont{7}{8.4}{\familydefault}{\mddefault}{\updefault}{\color[rgb]{0,0,0}$\kappa_{K2}=2.1788$}%
}}}}
\put(6751,-7111){\makebox(0,0)[lb]{\smash{{\SetFigFont{7}{8.4}{\familydefault}{\mddefault}{\updefault}{\color[rgb]{0,0,0}$\zeta_q$}%
}}}}
\put(2251,-6586){\makebox(0,0)[lb]{\smash{{\SetFigFont{7}{8.4}{\familydefault}{\mddefault}{\updefault}{\color[rgb]{0,0,0}$\kappa_{K1}=0.2$}%
}}}}
\put(8026,-8686){\makebox(0,0)[lb]{\smash{{\SetFigFont{7}{8.4}{\familydefault}{\mddefault}{\updefault}{\color[rgb]{0,0,0}$v_{K4}$}%
}}}}
\put(3376,-4111){\makebox(0,0)[lb]{\smash{{\SetFigFont{7}{8.4}{\familydefault}{\mddefault}{\updefault}{\color[rgb]{0,0,0}$u$}%
}}}}
\put(4726,-4111){\makebox(0,0)[lb]{\smash{{\SetFigFont{7}{8.4}{\familydefault}{\mddefault}{\updefault}{\color[rgb]{0,0,0}$y$}%
}}}}
\put(3451,-5086){\makebox(0,0)[lb]{\smash{{\SetFigFont{7}{8.4}{\familydefault}{\mddefault}{\updefault}{\color[rgb]{0,0,0}phase shift}%
}}}}
\thicklines
{\color[rgb]{0,0,0}\put(4201,-5761){\line(-1,-1){600}}
}%
\end{picture}%

\caption{Quantum-classical controller ($\Theta_K=\textrm{diag}(J,0_{2\times 2})$) for the plant of Figure \ref{fig:cavity4}.}
\label{fig:cavity5}
\end{center}
\end{figure}
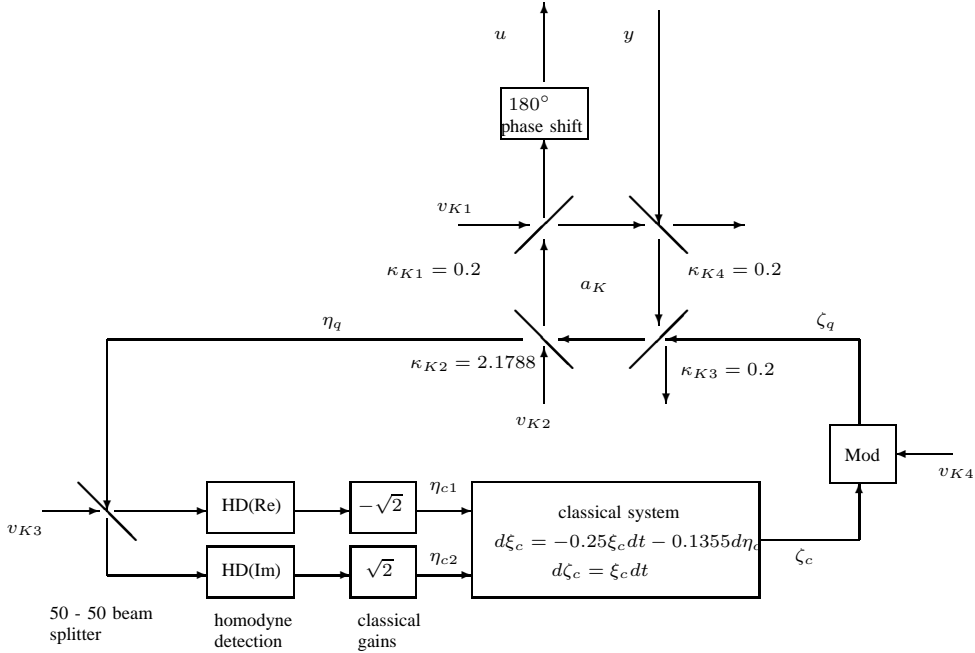

We choose $\Theta_K=\textrm{diag}(J,0_{2\times 2})$ in order to implement a degenerate canonical controller, with both classical and quantum degrees of freedom. We write $\xi=(\xi_q,\xi_c)^T$, where $\xi_q=(\xi_1,\xi_2)^T$ are classical and
$\xi_c=(\xi_3,\xi_4)^T$ are quantum variables.
A realization is shown in Figure \ref{fig:cavity5}, which consists of a four-mirror optical cavity, a classical system, and homodyne detection and modulation for interfacing the classical and quantum components.
The quantum noises in Figure \ref{fig:cavity5} are all canonical. The cavity has coupling coefficients $\kappa_{K1}=\kappa_{K3}=\kappa_{K4}=0.2$ and $\kappa_{K2}=1.33$. The interconnection fields are given by $d\eta_q=\xi_q dt+ dv_{K2}$, and $d\zeta_q=\zeta_c dt + dv_{K4}$, where $\eta_c=(\eta_{c1}, \eta_{c2})^T = (\eta_{q1}-v_{K31}, \eta_{q2}+v_{K32})^T$. For this realization we have
$$
 B_{K1}= \left[   \begin{array}{cccc}
 -0.4472 I & -1.4761 I & 0_{2 \times 2} & -0.4472I\\
 0_{2 \times 2} & -0.1355I & 0.1355\tilde I & 0_{2 \times 2}
 \end{array} \right]; \quad \tilde I=\left[\begin{array}{cc} 1 & 0 \\ 0 & -1 \end{array}\right]
 $$
 $$
 B_{K0}= \left[   \begin{array}{ccc}
-I & 0_{2 \times 2} & 0_{2 \times 2} \end{array} \right],
$$

\section{Conclusion}
\label{sec:conclusion}

In this paper we have formulated and solved an $H^\infty$ synthesis
problem for a class of non-commutative stochastic models. Models
important to quantum technology, such as those arising in quantum
optics, are included in this class. We have provided results for the physical realization of the controllers.
Our results are illustrated  with
examples from quantum optics, which demonstrate the synthesis of
quantum, classical and quantum-classical controllers.
Future work will include further development of the
approach initiated here, and application of the synthesis methods to
particular problems in quantum technology.


\useRomanappendicesfalse
\appendices
\section{Proofs}
\renewcommand{\thetheorem}{\thesection.\arabic{theorem}}

{\em Proof of Theorem \ref{thm:preserve}.} To preserve the
commutation relations for all $i,j=1,\ldots,n$ and all $t \geq 0$,
we must have $d[x_i,x_j]=0$ for all $i,j=1,\ldots,n$. We now
develop a general expression for $d[x_i,x_j]$. Indeed, let
$e_k=[\begin{array}{ccccccc} 0 & \ldots & 0 & 1 & 0 & \ldots & 0
\end{array}]\trp$, where the $1$ is in the $k$-th row.
It is easy to see
that for any $i,j \in \{1,\ldots,n\}$, $[x_i,x_j]=e_i\trp xx\trp
e_j-e_j \trp xx\trp e_i$.
Therefore, $d[x_i,x_j]=e_i\trp d(xx\trp) e_j-e_j\trp d(xx\trp)
e_i$.  Now, we expand
 $d(xx\trp)$ using the quantum Ito rule (e.g., see \cite{KRP92}) as follows:
\begin{eqnarray*}
d(xx\trp)&=&(dx) x\trp + x d(x\trp) +dx d(x\trp) \\
&=& Axx\trp dt + B dw x\trp + xx\trp A\trp dt + xd(w\trp)B\trp +
Axx\trp A\trp dt^2  \\
&& + Axd(w\trp)dt B\trp + B dw dt x \trp A\trp + B dw (dw)\trp B\trp
\\
&=& Axx\trp dt + B dw x\trp + xx\trp A\trp dt + xd(w\trp)B\trp +
B(dw)(dw)\trp B\trp.
\end{eqnarray*}
Substituting $dw=\beta_w dt + d\tilde{w}$ into the above  and
noting that $\beta_w \beta_w\trp dt^2$ and $\beta_w
d\tilde{w}\trp dt$ vanish to order $dt$ gives
\begin{eqnarray*}
d(xx\trp)&=& Axx\trp dt + B \beta_w x\trp dt  + B d\tilde{w} x\trp
+ xx\trp A\trp dt +x \beta_w\trp B \trp  dt+ x d\tilde{w}\trp B
\trp +  B d\tilde{w} d\tilde{w} \trp B\trp.
\end{eqnarray*}
We now write $A=[\begin{array}{cccc} A_1\trp & A_2\trp & \ldots &
A_n\trp
\end{array}]\trp$ and $B=[\begin{array}{cccc} B_{1}\trp & B_{2}\trp & \ldots
& B_{n}\trp \end{array}]\trp$, where the vectors $A_k$ and $B_{k}$
denote the $k$-th row of matrices $A$ and $B$, respectively. Then we have
\begin{eqnarray}
e_i\trp d(xx\trp) e_j&{=}& e_i\trp A xx\trp e_j dt +e_i\trp B\beta_w
x\trp e_j dt +e_i\trp Bd\tilde{w} x\trp e_j +
e_i\trp x x\trp A\trp e_j dt  \nonumber \\
&& + e_i x  \beta_w \trp B\trp e_j dt + e_i x d \tilde{w}\trp B\trp e_j + e_i\trp B d\tilde{w} (d\tilde{w})\trp B \trp e_j  \nonumber \\
&=& A_i x x_j dt + B_{i} \beta_w x_j dt + B_{i} d\tilde{w} x_j+
x_i A_j x dt + x_i B_{j} \beta_w dt + x_i B_{j} d\tilde{w}
+(B_{i} d\tilde{w}) (B_{j} d\tilde{w}). \label{eq_2}
\end{eqnarray}
Also we have
\begin{eqnarray}
e_j\trp d(xx\trp) e_i&=& A_j x x_i dt + B_{j} \beta_w x_i dt + B_{j} d\tilde{w} x_i + x_j A_i x dt + x_j B_{i} \beta_w dt
+ x_j B_{i} d\tilde{w} + 
(B_{j} d\tilde{w}) (B_{i}d\tilde{w}). \label{eq_3}\nonumber \\
\end{eqnarray}
Subtracting (\ref{eq_3}) from (\ref{eq_2}) gives us
\begin{eqnarray}
\label{eq_4}
e_i\trp d(xx\trp) e_j - e_j\trp d(xx\trp) e_i &=& ((A_i x) x_j
-x_j(A_i x))dt +  ((B_{i} \beta_w) x_j - x_j (B_{i} \beta_w))dt   \nonumber \\
&&+(B_{i} d\tilde{w}) x_j - x_j (B_{i} d\tilde{w}) + (x_i (A_j x) -
(A_j x)x_i)dt  \nonumber \\
&&+(x_i (B_{j} \beta_w) - (B_{j} \beta_w )x_i)dt + (x_i (B_{j} d\tilde{w}) -(B_{j}d\tilde{w})x_i)  \nonumber \\
&&+((B_{i} d\tilde{w}) (B_{j} d\tilde{w}) - (B_{j}
d\tilde{w})(B_{i} d\tilde{w})) \nonumber \\
&=& ((A_i x) x_j
-x_j(A_i x))dt +  ((B_{i} \beta_w) x_j - x_j (B_{i} \beta_w))dt   \nonumber \\
&&+(x_i (A_j x) -
(A_j x)x_i)dt  
+(x_i (B_{j} \beta_w) - (B_{j} \beta_w )x_i)dt  \nonumber \\
&&+((B_{i} d\tilde{w}) (B_{j} d\tilde{w}) - (B_{j}
d\tilde{w})(B_{i} d\tilde{w})).
\end{eqnarray}
Here we are using the fact that  elements of $d\tilde{w}$ commute
with those of $x$ and $\beta_w$ due to the adaptedness of $x$ and
$\beta_w$.
Hence,
\begin{eqnarray}
e_i\trp d(xx\trp) e_j-e_j\trp d(xx\trp) e_i &=& [A_i x,x_j]dt -
[x_j,B_{i} \beta_w] dt + [x_i, A_j x] dt +
[x_i,B_{j} \beta_w] dt + [B_{i} d\tilde{w}, B_{j} d\tilde{w}] \nonumber \\
&=& \sum_{k=1}^{n}A_{ik}[x_k,x_j]dt  -
\sum_{k=1}^{n}B_{ik}[x_j,\beta_{wk}]dt +
\sum_{k=1}^{n}A_{jk}[x_i,x_k]dt  \nonumber\\
&&+\sum_{k=1}^{n}B_{jk}[x_i,\beta_{wk}]dt +
\sum_{k=1}^{n}\sum_{l=1}^{n}B_{ik}B_{jl}
[d\tilde{w}_k,d\tilde{w}_l]
\nonumber\\
&=& \left(2i\sum_{k=1}^{n}A_{ik}\Theta_{kj} +
2i\sum_{k=1}^{n}A_{jk}\Theta_{ik} -\sum_{k=1}^{n}B_{ik}C^{x
\beta_w}_{jk} +
\sum_{k=1}^{n}B_{jk}C^{x\beta_w}_{ik}  \right. \nonumber\\
&&\left. +\sum_{k=1}^{n}\sum_{l=1}^{n}B_{ ik}B_{jl}
(F_{\tilde{w},kl}-F_{\tilde{w},lk})\right)dt, \label{eq_5}
\end{eqnarray}
where $C^{x\beta_w}_{ij}=[x_i,\beta_{wj}]$. Since
$C^{x\beta_w}{=}[C^{x\beta_w}_{ij}]_{i=1,\ldots,n,j=1\ldots,n_{\beta_w}}
= 0$ (by assumption) and $F_{\tilde w}-F_{\tilde w}\trp=2
T_{\tilde w}$, equation (\ref{eq_5}) takes the form
\begin{equation}
d( xx^T -  (xx^T)^T) = 2(iA\Theta+ i \Theta A^T + B T_{\tilde w}
B^T )dt \label{eq-thm2.2-6}
\end{equation}
from which the result follows.
\hfill $\Box$

{\em Proof of Theorem \ref{thm_phys_real}.} 

Let us first consider the case where $\Theta$ is canonical. If the
system is realizable then
(\ref{eq_coef_cond_A})-(\ref{eq_coef_cond_D}) holds. Since $U(t)$
is unitary for each $t \geq 0$, we have that
$d\left(x(t)x(t)\trp-(x(t)x(t)\trp)\trp\right)=0$; i.e., the
canonical commutation relations are preserved. By Theorem
\ref{thm:preserve} this is equivalent to
(\ref{eq_realize_cond_A}). Let $M_1,M_2,\ldots,M_{N_y}$ be column
vectors such that $[M_1 \smhsp M_2 \smhsp \ldots \smhsp
M_{N_y}]=\Lambda \trp
 [\begin{array}{cc} I_{N_y \times N_y} & 0
\end{array}]\trp$.  Then using (\ref{eq_coef_cond_B}) and
(\ref{eq_coef_cond_C}) we obtain the following after some
algebraic manipulations:
\begin{eqnarray*}
B[\begin{array}{cc} I_{n_y \times n_y} & 0_{n_y \times (n_w-n_y)}
\end{array} ]\trp &=& 2i\Theta [\smhsp -\Lambda^\dagger \smhsp \smhsp \Lambda\trp \smhsp]\Gamma
[\begin{array}{cc} I_{n_y \times N_y} & 0_{n_y \times (n_w-N_y)}
\end{array} ]\trp
\\&=&2\Theta [-\Im(M_1) \smhsp \Re(M_1) \smhsp \ldots \smhsp
-\Im(M_{N_y}) \smhsp \Re(M_{N_y})] \\
&=&\Theta \left(P_{N_y} \trp\left[ \begin{array}{cc} 0_{N_y \times N_y} & -I_{N_y \times N_y} \\
I_{N_y \times n_y} & 0_{N_y \times n_y}
\end{array}\right] P_{N_y} C\right)\trp \\
&=&\Theta C\trp P_{N_y}\trp \left[\begin{array}{cc} 0 & I_{N_y
\times N_y} \\ -I_{N_y \times N_y} & 0 \end{array}
\right]P_{N_y}\\
&=& \Theta C\trp \diag_{N_y}(J).
\end{eqnarray*}
Therefore, we conclude that (\ref{eq_realize_cond_B}),
(\ref{eq_realize_cond_A}) and (\ref{eq_coef_cond_D}) are necessary
for realizability.

Conversely, now suppose that (\ref{eq_realize_cond_B}),
(\ref{eq_realize_cond_A}) and (\ref{eq_coef_cond_D}) hold. We will
argue that these conditions are sufficient for realizability by
showing that they imply the existence a symmetric matrix $R$ and a
coupling matrix $\Lambda$ such that
(\ref{eq_coef_cond_A})-(\ref{eq_coef_cond_C}) are satisfied. First
we note that after some simple algebraic manipulation
 $ -i\Theta^{-1} B \Gamma^{-1}=i\Theta B \Gamma^{-1}=[-Z^\#
\smhsp Z],$ for some complex matrix $Z$. Hence $B=i\Theta[-Z^\#
\smhsp Z]\Gamma$. Substituting the last expression into
(\ref{eq_realize_cond_A}) and after further manipulations we get:
\[
i A\Theta + i\Theta A \trp-\frac{1}{2}\Theta (Z^\# Z\trp- Z
Z^\dagger)\Theta =0.
\]
Writing $Z^\# Z\trp-Z Z^\dagger =2i\Im(Z^\# Z\trp)$, we may
rewrite the last expression as follows:
\begin{eqnarray*}
iA\Theta + i\Theta A \trp-\frac{1}{2}\Theta(Z^\# Z\trp-Z Z^\dagger
)\Theta &=& iA\Theta
+ i\Theta A \trp-i\Theta\Im(Z^\# Z\trp)\Theta \\
&=& i\Theta(\Theta^{-1}A+A\trp\Theta^{-1}- \Im(Z^\# Z\trp))\Theta \\
&=& i\Theta(\Theta^{-1}A- (\Theta^{-1}A)\trp -\Im(Z^\# Z\trp))\Theta  \ = \ 0
\end{eqnarray*}
implying that $\Theta^{-1}A- (\Theta^{-1}A)\trp -\Im(Z^\#
Z\trp)=0$. Since $\Theta^{-1} A$ is real, we have the
decomposition $\Theta^{-1}A=-\Theta A=V+W$ for a unique pair of
real symmetric matrix $V$ and real skew symmetric matrix $W$ and
obtain the condition $2W-\Im(Z^\# Z\trp)=0$. Hence,
$W=\frac{1}{2}\Im(Z^\# Z\trp)$. Setting $R=\frac{1}{2}V$ and
$\Lambda=2Z\trp$, we get $A=2\Theta(R+\Im(\Lambda^\dagger
\Lambda))$ and $B=2i\Theta [-\Lambda^\dagger \smhsp \smhsp
\Lambda\trp]$ as desired, and also prove the second statement of
the theorem. After substituting the expression,  just obtained for
$B$ (in terms of $\Lambda$, $\Theta$, and $\Gamma$) into
(\ref{eq_realize_cond_B}) and more algebraic manipulations we then
get (\ref{eq_coef_cond_C}). Since the expression for $D$ has been
hypothesized as (\ref{eq_coef_cond_D}), we conclude that
(\ref{eq_realize_cond_B}), (\ref{eq_realize_cond_A}) along with
(\ref{eq_coef_cond_D}) gives matrices $A,B,C,D$ which are the
coefficients of a realizable system.

Now, we consider the case where $\Theta$ is degenerate canonical,
i.e., $\Theta=\diag(0_{n' \times n'},\diag_{\frac{n-n'
}{2}}(J))$. Let us write
\[
A=\left[ \begin{array}{cc} A_{11} & A_{12} \\ A_{21} & A_{22}
\end{array} \right] \quad B=\left[\begin{array}{cc} B_1 & B_2 \end{array} \right]
\quad C=\left[\begin{array}{cc} C_1 & C_2 \end{array} \right]
\]
with $A_{11} \in \rbb^{n' \times n'}$, $A_{12} \in \rbb^{n' \times
(n-n')}$, $A_{21} \in \rbb^{(n-n') \times n'}$, $A_{22} \in
\rbb^{(n-n') \times (n-n')}$, $B_1 \in \rbb^{n \times n_y}$, $B_2
\in \rbb^{n \times (n_w-n_y)}$, $C_1 \in \rbb^{n_y \times n'}$ and
$C_2 \in \rbb^{n_y \times (n-n')}$. Consider the following
augmentation:
\begin{eqnarray*}
d\tilde x(t)&=&\left[ \begin{array}{ccc} A_{11} & A_{12} & 0_{n'
\times n'} \\
A_{21} & A_{22} & 0_{(n-n') \times n'} \\
A'_1 & A'_2 & A''
\end{array} \right]\tilde x(t) dt +\left[ \begin{array}{cc} B_1 & B_2 \\ B'_1 & 0
\end{array}\right]dw(t)\\
d\tilde y(t)&=&\left[\begin{array}{cc} C & 0_{n_y \times n'}
\end{array} \right]\tilde x(t) dt + Ddw(t)
\end{eqnarray*}
where $B'_1=-C_1\trp P_{N_y}\trp \left[ \begin{array}{cc} 0 & I \\
-I & 0 \end{array} \right]P_{N_y}$, and $A_1'$, $A_2'$ and $A''$
satisfy the following:
\begin{eqnarray}
A'_1-(A'_1)\trp&=&i\left[ \begin{array}{cc} B'_1 & 0 \end{array} \right]T_w\left[\begin{array}{c} (B'_1)\trp \\
0 \end{array} \right] \nonumber\\
\left[\begin{array}{cc} A'' & -A'_2\diag_{\frac{n-n' }{2}}(J)
\end{array} \right]&=& -\left[\begin{array}{cc} A_{11}\trp &
A_{21}\trp \end{array} \right] - i\left[\begin{array}{cc} B'_1 & 0
\end{array} \right] T_w B\trp. \nonumber
\end{eqnarray}
It follows by inspection that such matrices $A_1'$, $A_2'$ and
$A''$ exist. Let $A'=[A'_1 \smhsp A'_2 ]$ and define
\[
\tilde A=\left[ \begin{array}{cc} A & 0_{n \times n'} \\ A' & A''
\end{array}\right] \quad \tilde B=\left[ \begin{array}{cc} B_1 & B_2 \\ B'_1 &
0
\end{array}\right] \quad  \tilde C=\left[ \begin{array}{cc} C & 0_{n \times n'}
\end{array}\right].
\]
If (\ref{eq_realize_cond_A}) holds then it can be  verified, by
direct substitution, that the matrices $\tilde A$ and $\tilde B$
satisfy:
\begin{equation}
\label{eq_aux_1}i\tilde A \tilde \Theta + i\tilde \Theta \tilde
A\trp + \tilde B T_w \tilde B\trp=0.
\end{equation}
Recalling  that $\tilde \Theta$ is only canonical up to
permutation,
 we now need to transform it into canonical form. To do this,
introduce the variable $z=P\tilde x$ where $P$ is a permutation
matrix such that $P\tilde \Theta P \trp=\diag_{\frac{\tilde
n}{2}}(J)$. Then the components of $z$ are a  relabelling of the
components of $\tilde x$. This gives us the following dynamics for
$z$:
\begin{eqnarray*}
dz(t)&=&P\tilde A P\trp z(t) + P\tilde B dw(t) \\
dy(t)&=&\tilde C P\trp z(t) dt + D dw(t).
\end{eqnarray*}
Denoting $\hat A=P\tilde A P\trp$, $\hat B=P\tilde B$, $\hat
C=\tilde C P\trp$, and $\hat \Theta=\diag_{\frac{\tilde n}{2}}(J)$
we see that (\ref{eq_aux_1}) implies that:
\begin{equation}
i\hat A \hat \Theta + i\hat \Theta \hat A\trp + \hat B T_w \hat
B\trp=0. \label{eq_aux_2}
\end{equation}
Continuing further using (\ref{eq_realize_cond_B}), we have the
following:
\begin{eqnarray}
\hat B\left[\begin{array}{c} I_{n_y \times n_y} \\
0_{(n_w-n_y) \times n_y}
\end{array} \right]&=&P\left[ \begin{array}{c} B \\ \begin{array}{cc} B'_1 & 0 \end{array} \end{array} \right]
\left[\begin{array}{c} I_{n_y \times n_y} \\
0_{(n_w-n_y) \times n_y}
\end{array} \right]\nonumber \\
&=& P\left[ \begin{array}{c} \Theta C\trp \\
-C_1\trp \end{array} \right]P_{N_y}\trp \left[ \begin{array}{cc} 0
& I \\ -I & 0 \end{array}
\right]P_{N_y} \nonumber \\
&=& P\tilde \Theta\left[ \begin{array}{c} C\trp \\
0\end{array} \right] P_{N_y}\trp \left[
\begin{array}{cc} 0 & I \\ -I & 0 \end{array} \right]P_{N_y} \nonumber \\
&=& (P\tilde \Theta P\trp) P\left[ \begin{array}{c} C\trp \\
0\end{array} \right] P_{N_y}\trp \left[
\begin{array}{cc} 0 & I \\ -I & 0 \end{array} \right] P_{N_y} \nonumber \\
&=& \hat \Theta \smhsp \hat C \trp P_{N_y}\trp \left[
\begin{array}{cc} 0 & I \\ -I & 0 \end{array} \right]P_{N_y}=\hat \Theta \hat C \trp \diag_{N_y}(J). \label{eq_aux_3}
\end{eqnarray}
If $D$ is given by (\ref{eq_coef_cond_D}) then (\ref{eq_aux_2})
and (\ref{eq_aux_3}) implies, as we have already shown for the
case of canonical $\Theta$, the system defined by the matrices
$(\hat A, \hat B, \hat C,D)$ is realizable in the sense of Point 1
of the theorem. Hence, the original system defined by the matrices
$(A,B,C,D)$ is then realizable in the sense of Point 2 of the
theorem.

Finally, suppose conversely that (\ref{linear-c}) is realizable
and let $(\tilde A, \tilde B, \tilde C, D)$ be a suitable
augmentation. Then \\ $(P\tilde AP\trp, P\tilde B, \tilde C P\trp,
D)$ is a quantum harmonic oscillator, with $P$ as defined before.
Hence, $P\tilde AP\trp$, $P\tilde B$, $\tilde C P\trp$, and $D$
are given by the right hand sides of
(\ref{eq_coef_cond_A})-(\ref{eq_coef_cond_D}) for a canonical
$\Theta$ and some $R$ and $\Lambda$. It follows that $\tilde A$,
$\tilde B$, $\tilde C$ and $\tilde D$ are given by the same set of
equations by replacing $\tilde \Theta$, $R$ and $\Lambda$ by
$\tilde \Theta=P\trp \Theta P$, $\tilde R=P\trp R P$ and $\tilde
\Lambda=\Lambda P$, respectively. We then have, from the same line
of arguments given for the case of canonical $\Theta$, that:
\begin{equation}
\tilde B\left[\begin{array}{c} I_{n_y \times n_y} \\
0_{(n_w-n_y) \times n_y}
\end{array} \right]=\tilde \Theta \tilde C\trp P_{N_y}\trp\left[\begin{array}{cc} 0 & I_{N_y \times N_y} \\ -I_{N_y \times N_y} & 0 \end{array}
\right]P_{N_y}=\tilde \Theta \tilde C\trp \diag_{N_y}(J),
\label{eq_aux_4}
\end{equation}
(\ref{eq_aux_1}) holds, and $D$ satisfies (\ref{eq_coef_cond_D}).
Reading off the first $n$ rows of both sides of (\ref{eq_aux_4})
then gives us (\ref{eq_realize_cond_B}), while reading of the
first $n$ rows and columns of both sides of (\ref{eq_aux_1}) gives
us (\ref{eq_realize_cond_A}), as required. This completes the
proof. \hfill $\Box$

%

The proof of  Theorem \ref{T1}  will use the following lemma.
\begin{lemma}
\label{L1}
Consider a real symmetric matrix $X$ and
corresponding operator valued
quadratic form $x^T Xx$ for the system (\ref{linear-diss}). Then the
following statements are equivalent:
\begin{enumerate}[(i)]
\item
There exists a constant $\lambda \geq 0$ such that
$
\langle\rho,x^T Xx\rangle \leq \lambda
$
 for all Gaussian states $\rho$.\footnote{Here
$\langle \rho, \cdot \rangle$ denotes the  expectation with
respect to the Gaussian state $\rho$.}
\item
The matrix $X$  is negative
semidefinite.
\end{enumerate}
\end{lemma}

{\em Proof} $(i) \Rightarrow (ii)$. To establish this part of the
lemma, consider a Gaussian state $\rho$ which has mean $\bar x$ and
covariance matrix $Y \geq 0$. Then, we can write
\begin{eqnarray}
\label{mean_variance_bound}
\langle\rho,x^T Xx\rangle &=& \sum_{i=1}^{n}\sum_{j=1}^{n}X_{ij}\langle\rho,x_i x_j\rangle \\
&=& \sum_{i=1}^{n}\sum_{j=1}^{n}X_{ij}[Y_{ij}+\bar x_i \bar
x_j]
\  =  \ \bar x^T X \bar x + \tr[XY].
\end{eqnarray}
Now for any constant $\alpha >  0$, consider the inequality
of part (i)  where $\rho$ is a Gaussian state with mean
$\alpha \bar x$ and
covariance matrix $Y$. Then it follows from this bound and
(\ref{mean_variance_bound}) that
$
\alpha^2 \bar x^T X \bar x + \tr[XY] \leq \lambda \
$
for all $\alpha >  0$. From this it immediately follows that $\bar x^T X
\bar x \leq 0$. However $\bar x$, the mean of the  Gaussian state
$\rho$ was arbitrary. Hence, we can conclude that condition $(ii)$ of
the lemma is satisfied.

 $(ii) \Rightarrow (i)$. Suppose that the matrix $X$ is negative
 semidefinite and let  $\rho$ be any Gaussian state
  and suppose that $\rho$ has mean $\bar x$ and
covariance matrix $Y \geq 0$. Then, it follows from
 (\ref{mean_variance_bound}) that
$
\langle\rho,x^T Xx\rangle = \bar x^T X \bar x + \tr[XY].
$
However, $X \leq 0$ and $Y \geq 0$ implies $\bar x^T X \bar x \leq 0$
and $\tr[XY] \leq 0$. Hence, $\langle\rho,x^T Xx\rangle \leq 0$ and condition $(ii)$
is satisfied with $\lambda = 0$.
\hfill $\Box$

{\em Proof of Theorem \ref{T1}.} Let the system be dissipative
with $V(x )=x^T  X x$. By Ito's rule, the table (\ref{ito-F}) and
the quantum stochastic differential equation (\ref{linear-diss})
we have
\begin{eqnarray}
d\langle V(x(t))  \rangle &=& \langle dx^T(t) X x(t) + x^T(t) X dx(t)
+ dx^T(t) X dx(t)  \rangle
 \nonumber  \\
&=& \left \langle 
x^T(t)({ A}^T X+X{ A}) x(t)
+ \beta_w^T(t) { B}^T  X x(t) + x^T(t) X { B} \beta_w(t)
+ \lambda_0  
\right \rangle dt ,
\label{calculation-ito-1}
\end{eqnarray}
where $\lambda_0$ is given by (\ref{equation-lambda}).
We now note that (e.g, see \cite[page 215]{KRP92})
$
\langle V(x(t) \rangle = \langle \rho, E_0 [ V(x(t)) ] \rangle ,
$
where $E_0$ denotes expectation with respect to $\phi$, and $\rho$ is an initial Gaussian state. Combining this with the integral of (\ref{calculation-ito-1}) and (\ref{diss}) we find that
\begin{eqnarray*}
\left\langle \rho,
   \int_0^t E_0 [
x^T(s)({ A}^T X+X{ A}) x(s)
+ \beta_w^T(s) { B}^T  X x(s) + x^T(s) X { B} \beta_w(s)
+ \lambda_0 + r(x(s),\beta_w(s)
] ds 
\right\rangle \leq \lambda t.
\end{eqnarray*}
Let $t \to 0$ to obtain
\begin{eqnarray*}
\left\langle \rho, x^T({ A}^TX+X{ A}) x + \beta_w^T{ B}^TX + x^T X{ B}\beta_w
+
\lambda_0
 + [x^T  \beta_w^T ] R \left[ \begin{array}{c} x \\ \beta_w \end{array} \right]  \right\rangle  \leq \lambda .
\end{eqnarray*}
Here, $x$ and $\beta_w$ denote the initial conditions. An application of
Lemma \ref{L1}   implies (\ref{MI-1}). Also, (\ref{MI-2}) is a
  straightforward consequence of this inequality when $R$ is replaced
  by $R+\epsilon I$ where $\epsilon > 0$.

To establish the converse part of the theorem, we first assume that
(\ref{MI-1}) is satisfied. Then with $V(x) = x^TXx$, it follows from
(\ref{calculation-ito-1}) that
\begin{eqnarray*}
\langle V(x(t))  \rangle - \langle V(x(0)) \rangle
 + \int_o^t\langle
r(x(s),\beta_w(s))\rangle ds
\leq   \lambda_0 t
\end{eqnarray*}
for all $t > 0$ and all $\beta_w(t)$. Hence, inequality (\ref{diss}) is
satisfied with $\lambda$ given by (\ref{equation-lambda}).

If matrix inequality (\ref{MI-2}) is satisfied, then it follows by
similar reasoning that there exists an $\epsilon > 0$ such that
\begin{eqnarray*}
\langle V(x(t))  \rangle - \langle V(x(0)) \rangle
+ \int_0^t\langle
r(x(s),\beta_w(s))+ \epsilon (x(s)^Tx(s) + \beta_w(s)^T\beta_w(s))\rangle ds
\leq    \lambda_0 t.
\end{eqnarray*}
Hence, inequality (\ref{diss}) is
satisfied with $\lambda=\lambda_0$ given by (\ref{equation-lambda}) and with
 $R$ replaced
by by $R+\epsilon I$.
\hfill $\Box$


{\em Proof of Theorem \ref{thm:H-synth}.}
Using the  Strict Bounded Real Lemma Corollary \ref{C3}, the  theorem
 follows directly from the corresponding classical $H^\infty$
result; e.g., see \cite{PAJ91,ZDG96,GL95}.
\hfill $\Box$

 The proof of
Theorem  \ref{thm_realization_hin} will use the following lemma.

\begin{lemma}
\label{lm_pos_shift}If $S$ is a Hermitian matrix then there is a
real constant $\alpha_0$ such that $\alpha I + S \geq 0$ for all
$\alpha \geq \alpha_0$.
\end{lemma}

{\em Proof.} Since $S$ is Hermitian it has real eigenvalues and is
diagonalizable. Hence $S=V^{\dagger}EV$ for some real diagonal
matrix $E$ and orthogonal matrix $V$. Now let $\alpha_0=-\lambda$,
where $\lambda$ is the smallest eigenvalue of $S$. The result
follows since $\alpha I + S=V^{\dagger}(\alpha I + E)V$ while
$\alpha I +E \geq 0$ for all $\alpha \geq \alpha_0$. \hfill $\Box$


{\em Proof of Lemma \ref{lemma_realization_hin}.}
%
The main idea is to explicitly construct matrices $R \in \rbb^{n_K
\times n_K}$, $\Lambda \in \cbb^{N_{v_K} \times n_K}$, $B_{K1} \in
\rbb^{n_K \times 2(N_{v_K} +N_y)}$ and $B_{K0} \in \rbb^{l_K
\times 2N_{v_K}}$, with $N_{v_K} \geq N_u$, such that
(\ref{eq_coef_cond_A})-(\ref{eq_coef_cond_D}) are satisfied by
identifying $A_K$, $B_K$, $C_K$, $[\begin{array}{cc} B_{K0} &
0_{l_K \times m_K}\end{array}]$, $\xi$, $w_K$ and $u$ with $A$,
$B$, $C$, $D$, $x$, $w$ and $y$, respectively. To this end, let
$Z=\frac{1}{2}\Theta_K^{-1}A=-\frac{1}{2}\Theta_K A$, with
$\Theta_K=\diag_{N_{\xi}}(J)$. We first construct matrices
$\Lambda_{b2}$, $\Lambda_{b1}$, $B_{K1,1}$ and $B_{K1,2}$
according to the following procedure:
\begin{enumerate}
\item Construct the matrix $\Lambda_{b2}$ according to (\ref{eq_b2_BK}).

\item Construct a real symmetric $n_K \times n_K$ matrix $\Xi_1$ such
  that the matrix
\begin{eqnarray*}
\Xi_2&=&\Xi_1+i\left(\frac{Z-Z\trp}{2}-\frac{1}{4}C_K\trp
P_{N_u}\trp \left[
\begin{array}{cc}  0 & I
\\ -I & 0 \end{array} \right]P_{N_u}C_K-\Im(\Lambda_{b2}^\dagger
\Lambda_{b2})\right)
\end{eqnarray*}
is non-negative definite. It follows from Lemma \ref{lm_pos_shift}
that such a matrix $\Xi_1$ always exists. \item Construct a matrix
$\Lambda_{b1}$ such that $\Lambda_{b1}^\dagger
\Lambda_{b1}=\Xi_2$, where $\Lambda_{b1}$ has {\em at least} $1$
row. This can be done, for example, using the singular value
decomposition of $\Xi_2$ (in this case $\Lambda_{b1}$ will have
$n_K$ rows).

\item Construct the matrices $B_{K1,1}$ and $B_{K1,2}$ according to
  equations (\ref{eq_def_BK11})
and (\ref{eq_def_BK12}), respectively.
\end{enumerate}

Let $R=\frac{1}{2}(Z+Z\trp)$. We now show that there exists an
integer $N_q^{v_K} \geq N_q^u$ such
conditions (\ref{eq_coef_cond_A})-(\ref{eq_coef_cond_D}) are
satisfied with the matrix $R$ as defined and with
$B_{K1}=[\begin{array}{cc} B_{K1,1} & B_{K1,2}
\end{array}]$ and
\begin{eqnarray} \Lambda=\left[ \begin{array}{c} \frac{1}{2}\left[
\begin{array}{cc} I & iI \end{array}\right]P_{N_u}C_k  \\
\Lambda_{b1} \\ \Lambda_{b2} \end{array} \right].
\label{eq_Lambda_aux}\end{eqnarray}

First note that necessarily $N_{v_K} \geq N_u+1 > N_u$ since
$B_{K1}$ has at least $2N_{u}+2$ columns. Also, by virtue of our
choice of $\Lambda_{b1}$ we have
\begin{eqnarray*}
\Im(\Lambda_{b1}^\dagger\Lambda_{b1})= \Im(\Xi_2)=
\frac{1}{2}(Z-Z\trp)-\frac{1}{4}C_K\trp P_{N_u}\trp \left[
\begin{array}{cc}  0 & I
\\ -I & 0 \end{array} \right]P_{N_u} C_K-\Im
(\Lambda_{b2}^\dagger \Lambda_{b2}),
\end{eqnarray*}
and hence
\begin{eqnarray*}
\Im(\Lambda^\dagger \Lambda) = \Im(\Lambda_{b1}^\dagger
\Lambda_{b1}) + \Im(\Lambda_{b2}^\dagger
\Lambda_{b2})+\frac{1}{4}C_K\trp P_{N_u}\trp \left[
\begin{array}{cc}  0 & I
\\ -I & 0 \end{array} \right]P_{N_u} C_K  = \frac{1}{2}(Z-Z\trp).
\end{eqnarray*}
Since $R=\frac{Z+Z\trp}{2}$, we have $R+\Im(\Lambda^\dagger
\Lambda)=Z$. Therefore, (\ref{eq_coef_cond_A}) is satisfied.

Now, as in the proof of Theorem \ref{thm_phys_real}, observe that
$
i\Theta_K B_K\diag_{N_y}(M^\dagger)P_{N_y}\trp=[\begin{array}{cc}
T & -T^\#
\end{array}] \
$
for some $n_K \times N_y$ complex matrix $T$. But by taking the
conjugate transpose of both sides of (\ref{eq_b2_BK}) which
defined $\Lambda_{b2}$, we conclude that
$T=-\Lambda_{b2}^\dagger$. Hence,
$
B_K=2i\Theta_K[\begin{array}{cc} -\Lambda_{b2}^{\dagger} &
\Lambda_{b2}\trp
\end{array} ]P_{N_y}\diag_{N_y}(M). \label{eq_BK_aux}
$
From (\ref{eq_def_BK11}) which defined $\Lambda_{b1}$,  we obtain
\begin{eqnarray}
B_{K1,1}&=&-i\Theta_K C_K \trp \diag_{N_u}(iJ) \nonumber \\
&=& -i\Theta_K C_K \trp
\diag_{N_u}(iJ)(2\diag_{N_u}(M^\dagger))\diag_{N_u}(M) \nonumber\\
&=& i\Theta_K C_K \trp \diag_{N_u}(\left[
\begin{array}{cc} -1 & 1 \\ i & i \end{array} \right])\diag_{N_u}(M) \nonumber \\
&=&i\Theta_K C_K\trp P_{N_u}\trp \left[\begin{array}{cc}-I & I
\\ iI & iI\end{array} \right]P_{N_u}\diag_{N_u}(M).
\label{eq_BK11_aux}
\end{eqnarray}

Combining (\ref{eq_def_BK12}), (\ref{eq_BK_aux}) and
(\ref{eq_BK11_aux}) gives us
\begin{eqnarray*}
\lefteqn{[\begin{array}{ccc}B_{K1,1} & B_{K1,2} &
B_{K}\end{array}]} \\
&=&2i\Theta_K \left[
\begin{array}{cc}
\frac{1}{2}C_K\trp P_{N_u}\trp \left[\begin{array}{cc} -I & I \\
iI & iI
\end{array} \right]P_{N_u} &
\left[\begin{array}{cc}-\Lambda_{b1}^\dagger & \Lambda_{b1}\trp
\end{array}  \right]P_{(N_{v_K}-N_u)}\end{array} 
\left[\begin{array}{cc} -\Lambda_{b2}^\dagger & \Lambda_{b2}\trp
\end{array}
\right]P_{N_y} \right]\\
&\quad& P_{N_{w_K}}\trp P_{N_{w_K}}\diag_{n_{w_K}}(M)\\
&=&2i\Theta_K \left[ \begin{array}{cccccc} -\frac{1}{2}C_K\trp
P_{N_u}\trp \left[\begin{array}{c} I \\ -iI
\end{array} \right] & -\Lambda_{b1}^\dagger &
-\Lambda_{b2}^\dagger & \frac{1}{2}C_K\trp P_{N_u}\trp
\left[\begin{array}{c} I \\ iI
\end{array} \right] & \Lambda_{b1}\trp & \Lambda_{b2}\trp
\end{array} \right] 
P_{N_{w_K}} \diag_{N_{w_K}}(M) \\
&=&2i\Theta_K \left[ \begin{array}{cccccc} -\frac{1}{2}C_K\trp
P_{N_u}\trp \left[\begin{array}{c} I \\ -iI
\end{array} \right] & -\Lambda_{b1}^\dagger &
-\Lambda_{b2}^\dagger & \frac{1}{2}C_K\trp P_{N_u}\trp
\left[\begin{array}{c} I \\ iI
\end{array} \right] & \Lambda_{b1}\trp & \Lambda_{b2}\trp
\end{array} \right]\Gamma \\
&=&2i\Theta_K \left[\begin{array}{cc} -\Lambda^\dagger & \Lambda\trp
\end{array} \right]\Gamma.
\end{eqnarray*}
Therefore, (\ref{eq_coef_cond_B}) is also satisfied. Moreover, it
is straightforward to verify (\ref{eq_coef_cond_C}) by
substituting $\Lambda$ as defined by
 (\ref{eq_Lambda_aux}) into the right hand side of
(\ref{eq_coef_cond_C}). Finally, since $N_{v_K} > N_u$, it follows
that $[\begin{array}{cc} B_{K0} & 0_{l_K \times m_k}
\end{array}]$ is precisely the right hand side of
(\ref{eq_coef_cond_D}). This completes the proof of Theorem
\ref{thm_realization_hin}. \hfill $\Box$

\noindent
{\em Proof of Lemma \ref{lem_MSS}}
We first observe that the system (\ref{cl2_Delta}) is mean square stable if and only
if it is  dissipative with a supply rate defined by the matrix $R=\mathrm{diag}(I,0)$.
 Hence, it follows from Theorem \ref{T1} that the system
(\ref{cl2_Delta}) is mean square stable if and only
if there  exists a real positive definite symmetric matrix
$X$ such that
$
\bar A^T X + X \bar A + I \leq 0.
$
Hence, using a standard Lyapunov result (e.g., see \cite{ZDG96}), it
follows that the system
(\ref{cl2_Delta}) is mean square stable if and only
if the matrix  $\bar A$ is asymptotically stable. \hfill $\Box$

{\em Proof of Theorem \ref{small_gain}}
It follows from Corollary \ref{C3} that the closed loop quantum
system (\ref{cl2}) is strictly bounded real with disturbance
attenuation $g$, then ${\tilde
  A}$ is a stable matrix and $\|{\tilde C}(sI-{\tilde A})^{-1}{\tilde
  B}+{\tilde D}\|_\infty <  g.$ From this, it follows using the
standard small gain theorem (e.g., see Theorem 9.1 on page 218 of
\cite{ZDG96}) that the matrix $\bar A =
\tilde A+ \tilde B \Delta  \tilde C$ is stable for all $\Delta$
satisfying (\ref{Delta_bound}). Hence using Lemma \ref{lem_MSS}, it
follows that  the true closed loop system (\ref{cl2_Delta}) is
mean square stable for all $\Delta$ satisfying (\ref{Delta_bound}).
\hfill $\Box$



\end{document}